\documentstyle[12pt]{article}
\documentstyle{fleqn}

\catcode`\@=11
\@addtoreset{equation}{section}
\def\theequation{\arabic{section}.\arabic{equation}}

\begin{document}
\begin{flushright}
OU-HET/183\\
hep-th/9311103\\
November 1993
\end{flushright}
\vspace{0.5in}
\begin{center}\Large{\bf Classical and Quantum Evolutions of the de Sitter
and the anti-de Sitter Universes in 2+1 dimensions }\\
\vspace{1cm}\renewcommand{\thefootnote}{\fnsymbol{footnote}}
\normalsize\ Kiyoshi Ezawa\footnote[1]{e-mail address: ezawa@oskth.kek.jp }
        \setcounter{footnote}{0}
\vspace{0.5in}

        Department of Physics \\
        Osaka University, Toyonaka, Osaka 560, Japan\\
\vspace{0.1in}
\end{center}
\vspace{1.2in}
\baselineskip 17pt
\begin{abstract}

 Two canonical formulations of the Einstein gravity in
2+1 dimensions, namely, the ADM formalism and the Chern-Simons gravity, are
investigated in the case of nonvanishing cosmological constant.
General arguments for reducing phase spaces of the two formalisms
are given when spatial hypersurface is compact.
In particular when the space has the topology of a sphere $S^{2}$ or
a torus $T^{2}$, the spacetimes constructed from these two formulations can
be identified and the classical equivalence between the ADM and the CSG is
shown. Moreover in the $g=1$ case the relations between their phase spaces,
and therefore between their quantizations, are given in almost the same
form as that
in the case when the cosmological constant vanishes. There are, however, some
modifications, the most remarkable one of which is that the phase space of
the CSG is in 1 to 2 correspondence with the one of the ADM when the
cosmological constant is negative.
\end{abstract}
\newpage

\baselineskip 20pt
\section{Introduction}

 \ \ \ \ For more than ten years, attention has been paid to the 2+1
dimensional gravity \cite{deser}\cite{deser2}\cite{marti}
as a useful toy model which
gives insights into the 3+1 dimensional quantum gravity. The pure Einstein
gravity in 2+1 dimensions is described by a finite number of global degrees of
freedom; no local degrees of freedom, the graviton, appear. This is because
Einstein's equations without matter tell us that the spacetime should be
locally diffeomorphic to a de Sitter, the Minkowski, or an anti-de Sitter
space according as the cosmological constant is positive, zero, or negative.
Dynamics in 2+1 dimensions is therefore much easier to deal with than that
in 3+1 dimensions. In addition, the 2+1
dimensional Einstein gravity shares with the 3+1 dimensional
gravity the propeties which lead to difficulties
in quantizing the latter. These are for example the facts that it is generally
covariant under diffeomorphisms and that it is not renormalizable when we use
metrics as basic variables. For these reasons we expect that the study of
the 2+1 dimensional quantum gravity yields some lessons in
quantizing the gravity in 3+1 dimensions.

It has been shown by Witten that the first order form of the  2+1 dimensional
Einstein gravity whose cosmological constant is positive, zero, or negative is
equivalent respectively to an $SO(3,1)$, an $ISO(2,1)$, or an $SO(2,2)$
gauge theory described by the Chern-Simons action \cite{witte}. We call it
"Chern-Simons gravity" (CSG). Many works have been done
about the CSG, which mainly focus on algebras of operators \cite{regge}
\cite{NRZ} \cite{stern} and on topological information
that can be extracted by path
integrals \cite{fujiw}\cite{carl3}.
A detailed analysis of the ADM formalism in 2+1 dimensions have been
given by Moncrief \cite{moncr} in the case of zero cosmological constant.
While the quantization is given when the genus of the time-slice
is one \cite{hosoy}, it is hard to get its physical interpretation.
Classical and
quantum relations between the CSG and the ADM are investigated by Carlip
\cite{carli} and others \cite{mess}\cite{ander} mainly in the torus case.
 From these works some geometrical aspects of the 2+1 dimensional quantum
gravity seem to be revealed.

The results obtained so far are mostly on the pure Einstein
gravity with zero cosmological constant, which describes the world
with no matter effects. What attracts our interest finally would be the
universe with some matter fields. One of the simplest ways to consider
matter effects is to introduce a cosmological term. However, there appear
to be relatively few works \cite{NRZ} which deal with the case with
nonvanishing cosmological constant from the viewpoint
of the canonical formalism.

In this paper, we investigate the pure Einstein gravity with nonzero
cosmological constant putting an emphasis  on geometrical aspects.
Our discussion
is based on the canonical formalism and therefore the topology of the
spacetime is restricted to $M\approx {\bf R}\times\Sigma$. We further assume
that the time-slice $\Sigma$ is a compact Riemann surface of genus $g$.
Moncrief's procedure for reducing the ADM formalism \cite{moncr} is extended
to the nonzero cosmological constant case. In the $g=0$ and $g=1$ cases we
explicitly observe that the spacetimes constructed by means of the CSG are
almost the same as those given by the ADM. The relations between the CSG and
the ADM are shown to be given in a way similar to that in the zero
cosmological constant case, with a few modifications. In particular for
$\Lambda<0$ and $g=1$, the correspondence between the phase spaces of the
CSG and of the ADM is shown to be 1 to 2. This is one of the main results in
this paper and gives an implication
that the CSG gives a "tunneling solution" similar to the one obtained by using
Ashtekar's formalism \cite{kodama}\cite{ashte}.

In \S 2 we apply Moncrief's method to the positive cosmological constant case,
and use the result to see a classical evolution of the torus
universe. In \S 3 after briefly reviewing the CSG, we look into the torus case
in detail. Based on the results of \S 2 and \S 3, classical and quantum
relations between the ADM formalism and the CSG are given in \S 4.
\S 5 is devoted to the study of special solutions in the $g=1$ case.
The $g=0$ case is investigated in \S 6.
Similar discussions are applicable to the anti-de Sitter case. We give
an outline for this case in \S 7.
In \S 8, after summarizing our results, we discuss remaining issues.

Here we give the conventions used in this paper:
\begin{enumerate}
 \item  $\mu,\nu,\rho,\cdots$ denote 2+1 dimensional spacetime indices and
   the metric $g_{\mu\nu}$ has the signature $(-,+,+)$.
 \item $i,j,k,\cdots$ are  used for spatial indices.
 \item $a,b,c,\cdots$ represent indices of the $SO(2,1)$ representation
   of the local Lorentz group, with the metric $\eta_{ab}={\rm diag}(-,+,+)$.
 \item $\hat{a},\hat{b},\hat{c},\cdots$ denote indices of the $SO(3,1)$ (or
   $SO(2,2)$) vector representation of the (anti-)de Sitter group.
   The metric is given by $\eta_{\hat{a}\hat{b}}={\rm diag}(-,+,+,+)$ in the
   de Sitter case, and by $\eta_{\hat{a}\hat{b}}={\rm diag}(-,+,+,-)$ in the
   anti-de Sitter case.
 \item $\alpha,\beta,\gamma,\cdots$ mean indices of the $SO(3)$ subgroup.
 \item $A,B,C,\cdots(\bar{A},\bar{B},\bar{C},\cdots)$ are indices of the
   $SL(2,{\bf C})$
   (anti-)self-dual spinor representation of the de Sitter group.
 \item $\epsilon_{abc}$ is the totally antisymmetric pseudo-tensor with
   $\epsilon_{012}=-\epsilon^{012}=1$.
 \item $\epsilon^{\mu\nu\rho}$ denotes the totally antisymmetric tensor
   density with $\epsilon^{t12}=-1$.
\end{enumerate}


\def\lap{\bigtriangleup}

\section{The reduced ADM formalism }

 \ \ \ \  Two reductions of the 2+1 dimensional ADM formalism with vanishing
cosmological constant are formulated
by Moncrief \cite{moncr} and Hosoya and Nakao \cite{hosoy}  \cite{nakao}.
In this section, we apply Moncrief's analysis to the case with positive
cosmological constant.
Particularly in the case of genus 1 ( $\Sigma\approx T^{2}$ ), we explicitly
solve the reduced Hamilton equations and construct a spacetime
from the classical solution. The resulting spacetime turns out to be the
quotient space of a covering space of the 3-dimensional de Sitter space
modulo a subgroup of the $SO(3,1)$ group.

\subsection{The general consideration }

 \ \ \ \ Our starting point is the Einstein action with a positive
cosmological constant $\Lambda$:
\begin{equation}
I_{E} = \int_{M} d^{3}x \sqrt{-^{(3)}g}(^{(3)}R-2\Lambda) ,\label{eq:E.ac}
\end{equation}
where $M$ denotes a 2+1 dimensional spacetime and the superscript $^{(3)}$
means that the quantity is defined on $M$. \\
In order to proceed to the ADM formalism, we restrict $M$ to be
of the topology $R\times\Sigma$ and express the spacetime metric as
$$
ds^{2}=^{(3)}g_{\mu\nu}dx^{\mu}dx^{\nu}
      =-N^{2}dt^{2}+g_{ij}(dx^{i}+N^{i}dt)(dx^{j}+N^{j}dt),
$$
where $g_{ij}$ is the induced metric on $\Sigma$, $N$ and $N^{i}$ are called
the lapse function and the shift vector respectively. Using this 2+1
decomposition, the action (\ref{eq:E.ac}) is rewritten as follows:
\begin{equation}
I_{ADM} = \int dt\int_{\Sigma}d^{2}x(\pi^{ij}\dot{g}_{ij}-N^{i}{\cal H}_{i}-
N{\cal H}), \label{eq:action}
\end{equation}
where
$$
\pi^{ij}\equiv \sqrt{g}(K^{ij}-g^{ij}K) \mbox{ \ , \ }\left(
K_{ij}\equiv\frac{1}{2N}(\dot{g}_{ij}-\nabla_{i}N_{j}-\nabla_{j}N_{i})\right)
$$
is the momentum conjugate to $g_{ij}$,
\begin{equation}
{\cal H}_{i}=-2\nabla_{j}\pi^{j}_{i}
\end{equation}
 and
\begin{equation}
{\cal H}= \frac{1}{\sqrt{g}}(\pi^{ij}\pi_{ij}-(\pi^{i}_{i})^{2})
          -\sqrt{g}(R-2\Lambda)
\end{equation}
are respectively called the momentum and the Hamiltonian constraints.\\
As is well known, this system is a 1st class constraint system and in
2+1 dimensions there remain no local degrees of freedom.
We will reduce this action to a canonical system of a finite number of
global degrees of freedom. More precisely we will reduce the phase space
to the cotangent bundle of the Teichm\"{u}ller space
${\bf T}^{\ast}{\cal T}(\Sigma)$
 of $\Sigma$, by following Moncrief's method.

First we solve the constraints.
We employ the fact that any smooth metric $g_{ij}$ on $\Sigma$ is globally
conformal to a constant curvature metric $h_{ij}$ \cite{aubin}\cite{fish}:
\begin{equation}
g_{ij}=e^{2\lambda}h_{ij} \mbox{ , \ }
   R(h)=\left\{ \begin{array}{rll}
          1 & for & g=0 \\
		  0 & for & g=1 \\
		 -1 & for & g\geq 2,
		 \end{array} \right. \label{eq:conf}
\end{equation}
and we decompose the momentum $\pi^{ij}$ as follows:
\begin{equation}\begin{array}{lcl}
\pi^{ij} &=& \pi^{ij TT}+e^{-2\lambda}\sqrt{g}
           (\nabla^{i}Y^{j}+\nabla^{j}Y^{i}-g^{ij}\nabla_{k}Y^{k})
		   +\frac{1}{2}\tau\sqrt{g}g^{ij} \\
		 &=& e^{-2\lambda}p^{ij}+e^{-2\lambda}\sqrt{h}
		   (\tilde{\nabla}^{i}Y^{j}+\tilde{\nabla}^{j}Y^{i}
		   -h^{ij}\tilde{\nabla}_{k}Y^{k})
		   +\frac{1}{2}\tau\sqrt{h}h^{ij},
	\end{array} \label{eq:ort.dec.}
\end{equation}
where $\tau\equiv g_{ij}\pi^{ij}/\sqrt{g}$ is called the mean curvature, and
$\pi^{ijTT}$ and $p^{ij}$ represent the transeverse-traceless parts of
$\pi^{ij}$  with respect to the metric $g_{ij}$ and $h_{ij}$ respectively.
The tilde denotes the operation associated with  $h_{ij}$.

Let us look into the momentum constraint. To simplify the analysis, we
impose a particular slicing condition called York's time slice:
$$
\tau= \mbox{constant on }\Sigma.
$$
Then the momentum constraint ${\cal H}_{i}=0$ implies that the vector $Y^{i}$
be a conformal killing field with respect to $h_{ij}$ and can be dropped from
 the decomposition (\ref{eq:ort.dec.}). As for $p^{ij}$, we know that:
(i) $p^{ij}\equiv0$ for $g=0$;
(ii) $p^{ij}$ is covariantly constant relative to $h_{ij}$ and forms the
2 dimensional space for $g=1$; and
(iii) the space of $p^{ij}$'s is $6g-6$ dimensional for $g\geq 2$.

Now we look for the solution to the Hamiltonian constraint. \\
By using eqs.(\ref{eq:conf})(\ref{eq:ort.dec.}) with $Y^{i}=0$ and
the conformal identity
$$
R=e^{-2\lambda}[R(h)-2\tilde{\lap}\lambda],
$$
the Hamiltonian constraint ${\cal H}=0$ is reduced to the following nonlinear
elliptic equation for the conformal factor $\lambda$
\begin{equation}
\tilde{\lap}\lambda\equiv
\frac{1}{\sqrt{h}}\partial_{i}(\sqrt{h}h^{ij}\partial_{j}\lambda)
=Ae^{2\lambda}-Be^{-2\lambda} +C,\label{eq:ellip}
\end{equation}
where
\begin{equation}
\left\{ \begin{array}{lcc}
  A &=& \frac{1}{4}\tau^{2}-\Lambda \mbox{ \ }(= const. on \Sigma)\\
    & &                                                          \\
  B &=& \frac{h_{ij}h_{kl}p^{ik}p^{jl}}{2h} \\
    & &                                      \\
  C &=& \frac{1}{2}R(h).
  \end{array} \right.
\end{equation}
We should bear in mind that $B$ is always non-negative. Making use of
Moncrief's "existence and uniqueness theorem"\cite{moncr}
\cite{moncr2}, we have obtained the following result\footnote{
We assume that $h_{ij}$ and $p^{ij}$ are $C^{\infty}$.}:
\begin{enumerate}
\item for $g\geq2$ ( $C=-1/2$ ),
 \begin{enumerate}
  \item no solutions exist if $ \tau^{2}-4\Lambda\leq0 $,
  \item a unique solution $ \lambda\in C^{\infty} $ always exists if
  $ \tau^{2}-4\Lambda>0 $.
 \end{enumerate}
\item for $g=1$ ( $C=0$, $B=const.on \Sigma$ ),
 \begin{enumerate}
  \item no solutions exist if $ \tau^{2}-4\Lambda<0 $,
  \item ( $p^{ij}=0$, $\lambda=$ an arbitrary constant on $\Sigma$ ) is a
  solution if $ \tau^{2}-4\Lambda=0 $,
  \item if $ \tau^{2}-4\Lambda>0 $, a unique solution
  $$
  e^{2\lambda}=\sqrt{\frac{2h_{ij}h_{kl}p^{ik}p^{jl}}{(\tau^{2}-4\Lambda)h}}
  $$
  exists unless $ p^{ij}=0 $.
 \end{enumerate}
\item for $g=0$ ( $C=1/2$, $B=0$ ),
 \begin{enumerate}
 \item there exists a 3-parameter family of solutions if
 $\tau^{2}-4\Lambda<0$,\footnote{Moncrief's discussion for the uniqueness of
 the solution (the appendix of \cite{moncr2}) cannot be applied
 in this case because
 $$ \frac{\partial}{\partial\lambda}\{(r.h.s \mbox{ }of\mbox{ }
 (\ref{eq:ellip})\}<0. $$ }
 \item no solutions exist if $ \tau^{2}-4\Lambda\geq0 $.
 \end{enumerate}
\end{enumerate}

Next we consider the ADM action. The reduced action is obtained by the
procedure parallel to that of Moncrief \cite{moncr} and we finally find:
\begin{equation}
I^{\ast\ast}_{ADM}=\int dt \{p^{\alpha}\dot{m}_{\alpha}-H(m_{\alpha},
                                                      p^{\beta},\tau)\},
	\label{eq:ADMf}
\end{equation}
where
$$
m_{\alpha}\quad(\alpha=1,\cdots,n_{g})\quad, \quad
 n_{g}=\left\{\begin{array}{ccl}
          2 & \mbox{ \ for }& g=1 \\
		  6g-6 & \mbox{ \ for }& g\geq2 . \end{array} \right.
$$
are the Teichm\"{u}ller parameters and $p^{\alpha}$'s are their conjugate
momenta. The Hamiltonian of this reduced
system is given by
$$
H(m_{\alpha},p^{\beta},\tau)\equiv \frac{d\tau}{dt}
                           \int_{\Sigma}d^{2}x e^{2\lambda}\sqrt{h},
$$
where $\lambda$ is the solution of eq.(\ref{eq:ellip}). We should notice
that $H$ is proportional to the area of the time-slice $\Sigma$ and,
on a given $\Sigma$, the $p^{\alpha}$'s and $H$ are
independent of the particular chosen cross section, namely the spatial gauge
fixing.

By solving this Hamiltonian system, we can deduce the classical evolution of
$(h_{ij},p^{ij})$ and hence that of $(g_{ij},\pi^{ij})$. To determine a
spacetime metric, however, the lapse $N$ and the shift $N^{i}$ remain to be
specified. In the framework of the classical theory, once we have fixed a
gauge, we can determine the lapse and the shift from equations of motion
in the original system (\ref{eq:action}).
The detailed analysis is given in ref.\cite{moncr} when the cosmological
constant is zero. Here we summarize briefly the procedure and the result
in the case of positive cosmological constant.

The equation for the lapse $N$ is obtained from the time derivative of the
mean curvature
$$
\frac{\partial\tau}{\partial t}=\{ \tau, \int_{\Sigma} d^{2}x(N^{i}{\cal H}_{i}
                                  +N{\cal H}) \}_{P.B.},
$$
and it is reweitten as
\begin{equation}
e^{2\lambda}\frac{\partial\tau}{\partial t}=
-\tilde{\lap}N+N\{ e^{-2\lambda}\frac{h_{ij}h_{kl}p^{ik}p^{jl}}{h}
                  +e^{2\lambda}\frac{\tau^{2}-4\Lambda}{2} \}
			\equiv  -\tilde{\lap}N+Nq.
			    \label{eq:lapseD}
\end{equation}
As for its solution, the result depends on the genus $g$ of $\Sigma$.
\begin{enumerate}
 \item for $g\geq2$, we are interested in the case $ \tau^{2}-4\Lambda>0 $.
  Since $q$ is positive everywhere, the discussion in the appendix of ref.
  \cite{moncr2} holds and the condition $\tau=t$ yields a unique solution
  for $N$ .
 \item for $g=1$, we have to consider the following two cases:
  \begin{enumerate}
   \item if $ \tau^{2}-4\Lambda=0 $, the preceding result tells us that
   $p^{ij}\equiv 0$ and that $\lambda$ should be an arbitrary constant on
   $\Sigma$. Then the above equation becomes the Laplace equation for $N$ and
   forces $N$ to be a constant on $\Sigma$.
   \item if $ \tau^{2}-4\Lambda>0 $, $\tau$ have to be a monotonic function of
   the time $t$ in order that the metric should not be singular. Equation
   (\ref{eq:lapseD}) has a unique solution which is
   \begin{equation}
   N=\frac{\partial\tau}{\partial t}\cdot \frac{1}{\tau^{2}-4\Lambda}=
      const. on\mbox{ \ }\Sigma. \label{eq:lapseT}
   \end{equation}
  \end{enumerate}
 \item for $g=0$, in contrast to $g\geq1$, only the case in which
 $ \tau^{2}-4\Lambda<0 $ requires our investigation. In this case $q$ is
 negative on $\Sigma$ and so equation (\ref{eq:lapseD}) has in general
 many solutions.
\end{enumerate}
The shift vector is determined by using the equation of motion of the tensor
density $ \sqrt{h}h^{ij}=\sqrt{g}g^{ij} $. We refer the detailed argument to
\cite{moncr} and list the results:
\begin{enumerate}
\item for $g\geq2$, the shift $N^{i}$ is uniquely determined since there is no
conformal killing vector on $\Sigma$.
\item for $g=1$, if we choose a constant metric $h_{ij}$ on $\Sigma$
as a cross section, $N^{i}$ is a time dependent linear combination of two
killing vectors, which are the conformal killing vectors.
\item for $g=0$, if we choose as $h_{ij}$ a standard metric on $S^{2}$
$$
d\sigma^{2}=2(d\theta^{2}+\sin^{2}\theta d\phi^{2}),
$$
then $N^{i}$ is a time dependent linear combination of six conformal
killing vectors.
\end{enumerate}

\subsection{The classical evolution of the torus $T^{2}$}

\ \ \ In the last subsection, we have obtained the reduced action of the 2+1
dimensional Einstein gravity with positive cosmological constant.
The lapse function and the shift vector have been given implicitly.
In principle, we can construct spacetimes which are solutions of the
Einstein equations by solving the Hamilton equations given by the action
(\ref{eq:ADMf}) and by choosing an appropriate cross section of the bundle
$\pi:\{ h_{ij}\}\rightarrow {\cal T}(\Sigma) $\footnote{The other
elements required to construct spacetimes are
determined once numerical values of $(h_{ij},p^{kl})$ are fixed. For
example, $\lambda$ and $\tau$ are given by equation(\ref{eq:ellip}) and
the temporal coordinate condition respectively. }.\\
However in practice, we meet various obstructions to carry out this program.
Indeed for the $g\geq2$ case, where we know that a unique solution $\lambda$
for eq.(\ref{eq:ellip}) exists under the proper condition, the explicit form
of the solution $\lambda$ and hence that of the Hamiltonian $H$ are not known
\cite{moncr}. It seems almost impossible to give a
time evolution  of the Teichm\"{u}ller parameters of the Riemann surface
with $g\geq2$ .
It appears hard to address the $g=0$ case also because the equations
have many solutions. We will deal with this case in \S 6.

Compared to the above two cases, the $g=1$ case is rather tractable. In this
case the solution to (\ref{eq:ellip}) is unique and its form is explicitly
determined. Dynamics of the canonical variables is therefore explicitly
given and the spacetime can indeed be constructed. In this subsection,
we investigate this case in detail. The resulting spacetime turns out to be
the quotient space of (a covering space of) a de Sitter space $dS^{3}$,
modulo a subgroup of its isometry $SO(3,1)$.

We know the reduced action for $\Sigma\approx T^{2}$:
\begin{equation}
I^{**}_{ADM}=\int dt(p_{1}\dot{m}_{1}+p_{2}\dot{m}_{2}-\frac{d\tau}{dt}
              \int d^{2}x e^{2\lambda}\sqrt{h}).
\label{eq:ADMaction}
\end{equation}
We choose as a cross section a constant metric $h_{ij}$ on $\Sigma$ with
$h=1$. We can express this metric using the Teichm\"{u}ller parameters as
\begin{equation}
h_{ij}dx^{i}dx^{j}=\frac{1}{m_{2}}\{ (dx+m_{1}dy)^{2}+(m_{2}dy)^{2}\},
\end{equation}
where $(x^{1},x^{2})\equiv(x,y)$ are coordinates on $T^{2}$ with period 1.
Now we will consider the case $\tau^{2}>4\Lambda$. The case
$\tau^{2}=4\Lambda$ will be discussed in \S 5. We impose the following
temporal coordinate condition
\begin{equation}
\tau=2\sqrt{\Lambda}\coth(2t),\qquad (\mbox{ } t>0\mbox{ }).\label{eq:tgauge}
\end{equation}
Then the Hamiltonian is written as
\begin{equation}
H\equiv \frac{d\tau}{dt}\int d^{2}x e^{2\lambda}\sqrt{h}
        =\frac{-1}{\sinh t\cosh t}m_{2}\sqrt{(p_{1})^{2}+(p_{2})^{2}}.
\end{equation}
The time coordinate $t$ is suitable to see geometrical aspects of the
solution. The solution of the Hamilton equations
\begin{equation}
\frac{d}{d\tau}(m_{\alpha},p_{\alpha})=\{(m_{\alpha},p_{\alpha}),H\}_{P.B.}
\qquad(\alpha=1,2),
\end{equation}
is
\begin{eqnarray}
m_{1}&=&\frac{uv+\alpha\beta\coth^{-2}t}{u^{2}+\alpha^{2}\coth^{-2}t},
\nonumber \\*
m_{2}&=&\frac{(u\beta-v\alpha)\coth^{-1}t}{u^{2}+\alpha^{2}\coth^{-2}t},
\nonumber \\*
p_{1}&=&-\frac{2}{\sqrt{\Lambda}}\alpha u, \qquad and \nonumber \\*
p_{2}&=&-\frac{\coth t}{\sqrt{\Lambda}}(u^{2}-\alpha^{2}\coth^{-2}t).
		\label{eq:phase space}
\end{eqnarray}
where $\alpha$, $\beta$, $u$, and $v$ are the constants of motion.
As in the case without a cosmological term, the point $(m_{1},m_{2})$
in the Teichm\"{u}ller space moves on a
geodesic of the Poincar\'{e} metric :
$$ \frac{1}{m_{2}^{\mbox{ }2}} (dm_{1}^{\mbox{ }2}+dm_{2}^{\mbox{ }2}), $$
which does not form a complete semicircle because $\coth t$ ranges
from 1 to $\infty$ (Figure 1).

Let us now investigate the spacetime geometry constructed from the
classical solution (\ref{eq:phase space}). Since we have fixed the time
coordinate $t$, the lapse is determined by substituting (\ref{eq:tgauge}) into
eq.(\ref{eq:lapseT})
\begin{equation}
N=\frac{-1}{\sqrt{\Lambda}}.
\end{equation}
The result of the last subsection tells us that the shift $N^{i}$ is a
linear combination of the two conformal killing vectors, which can be absorbed
by a redefinition of the origin of each time-slice. We set
\begin{equation}
N^{i}=0.\label{eq:shift}
\end{equation}
The conformal factor $\lambda$ is given by
$$
e^{2\lambda}=\frac{m_{2}\sqrt{(p^{1})^{2}+(p^{2})^{2}}}
                 {\sqrt{\tau^{2}-4\Lambda}}
			=\frac{\sinh t\cosh t}{\Lambda}(u\beta-v\alpha) .
$$

By putting these results together, we see that the spacetime
is described by the metric
\begin{eqnarray}
ds^{2}&=&-(Nd\tau)^{2}+e^{2\lambda}\frac{1}{m_{2}}
         \{dx^{2}+ 2m_{1}dxdy+ (m_{1}^{2}+m_{2}^{2})dy^{2}\}
\nonumber \\*
      &=&\frac{1}{\Lambda}[-dt^{2}+\cosh^{2}t\mbox{ }d\varphi^{2}
	                      +\sinh^{2}t\mbox{ }d\theta^{2}],
\label{eq:metric}
\end{eqnarray}
accompanied by the periodicity condition
\begin{equation}
(\varphi,\theta)\sim(\varphi+u,\theta+\alpha)\sim(\varphi+v,\theta+\beta).
\label{eq:period}
\end{equation}
Here we have transformed the coordinates as
\begin{equation}\left\{\begin{array}{l}
\varphi=ux+vy\\
\theta=\alpha x+\beta y  \quad. \end{array}\right.
\end{equation}

We can embed this spacetime into the 3 dimensional de Sitter space $dS^{3}$
and hence into the 3+1 dimensional Minkowski space $M^{3+1}$.
For this purpose, we parametrize the points on $dS^{3}$ as follows:
\begin{equation}
(T,X,Y,Z)=\frac{1}{\sqrt{\Lambda}}(\sinh t\cosh\theta,\sinh t\sinh\theta,
                   \cosh t\cos\varphi,\cosh t\sin\varphi).
				   \label{eq:embed}
\end{equation}
These points satisfy
$$ \begin{array}{l}
-T^{2}+X^{2}+Y^{2}+Z^{2}=\frac{1}{\Lambda},  \\ \quad
T>|X|\mbox{ \ } (Y^{2}+Z^{2}>\frac{1}{\Lambda}),\end{array}
$$
which imply that they are in a particular subspace of $dS^{3}$ (Figure 2a).
The de Sitter metric induced from the Minkowski metric
$$ ds^{2}=-dT^{2}+dX^{2}+dY^{2}+dZ^{2} $$
under the parametrization (\ref{eq:embed}) reproduces the r.h.s. of eq.(
\ref{eq:metric}) as is expected. \\
The periodicity condition (\ref{eq:period})
is expressed by the action of two isometries of $dS^{3}$:
\begin{equation}
^{t}(T,X,Y,Z)\sim E_{1}\cdot^{t}(T,X,Y,Z)\sim E_{2}\cdot^{t}(T,X,Y,Z),
\label{eq:period2}
\end{equation}
where
\begin{equation}\begin{array}{l}
E_{1}=\left( \begin{array}{cccc}
        \cosh\alpha & \sinh\alpha & 0 & 0 \\
		\sinh\alpha & \cosh\alpha & 0 & 0 \\
		0 & 0 &     \cos u & -\sin u \\
		0 & 0 &     \sin u & \cos u
		\end{array}\right) ,\\
E_{2}=\left( \begin{array}{cccc}
        \cosh\beta & \sinh\beta & 0 & 0 \\
		\sinh\beta & \cosh\beta & 0 & 0 \\
		0 & 0 &     \cos v & -\sin v \\
		0 & 0 &     \sin v & \cos v
		\end{array}\right).
  \end{array}	\label{eq:SO31}
\end{equation}
It is not considered, however, that the above "embedding"
works in a exact sense. Compare the two
periodicity conditions (\ref{eq:period}) and (\ref{eq:period2}).
While the "embedded" condition (\ref{eq:period2}) identifies $u+2\pi$ and
$v+2\pi$ with $u$ and $v$ respectively, such identification does not appear in
the original periodicity (\ref{eq:period}). It is therefore natural to
consider that the spacetime we have obtained by  means of ADM is
actually the quotient
space of {\em the universal covering space of} the de Sitter space $dS^{3}$:
(\ref{eq:embed}) with  $t>0$, $\theta\in(-\infty,\infty)$, and
$\varphi\in(-\infty,\infty)$, modulo the identification (\ref{eq:period})
\footnote{To distinguish the points with $\varphi$ and those with
$\varphi+2n\pi (n\in{\bf Z})$,
we have to cut the submanifold(\ref{eq:embed}) of
$dS^{3}$ along e.g. the surface $\varphi=0$, patch its infinitely many
copies along that surface, and form the universal
covering of the submanifold.}.
This identification belongs to the universal covering
$\widetilde{SO(3,1)}$ of the isometry group of $dS^{3}$.

Such prescription is indeed possible. If we consider the whole de Sitter space
as in ref.\cite{deser2}, the similar identification occurs when the antipodal
singularities exist on the time-slice which originally have the topology of a
sphere $S^{2}$. In our case, since the topology of the time-slice
of the primitive space\footnote{We use the term "primitive space" to mean the
prototype of the "fundamental region", which is the spacetime before
identification. The fundamental region is a covering space of the
primitive space.} is a cylinder $R\times S^{1}$ (Fig.2b),
such singularities are prepared at the outset. Thus we have
no problem in carrying out the operation mentioned above.

There is another advantage in using the universal covering space.
For illustration we consider identifying by $E_{1}$ with $\alpha=0$, which
represents a pure rotation in the $(Y,Z)$-plane.
If we use as a fundamental region the de Sitter space(\ref{eq:embed}) as it
is, then the result of the identification by $E_{1}$ generically,
more precisely when $u/\pi$ is irrational, collapses to a singularity
since the $\varphi$-direction originally forms a circle $S^{1}$.
On the other hand, use of the universal
covering avoids such a singularity because the $\varphi$-
direction originally has the topology $R^{1}$.

To summarize, as a result of solving the Hamilton equations of the reduced ADM
formalism in the case of $\Sigma\approx T^{2}$, we constructed the spacetime.
It is the quotient space:
\begin{equation}
M={\cal F}/G,
\end{equation}
where ${\cal F}$ is the fundamental region, which is the
universal covering of the
subspace (\ref{eq:embed}) of the de Sitter space $dS^{3}$, and $G$ is the
group generated by two identifications in (\ref{eq:period}). Note that this
$G$ is a subgroup of the universal covering of the de Sitter group
$\widetilde{SO(3,1)}$.

Finally we make a few remarks on the geometrical feature of the time-slice:
i) it is intrinsically flat because the time-slice of the primitive space
has the geometry of a cylinder $S^{1}\times R^{1}$ with all $S^{1}$ having the
same radius; and ii) its evolution is considerably different from that in the
$\Lambda=0$ case. The time-slice expands exponentially from a wire-like
singularity while twisting and its shape approaches to a {\em nonsingular}
torus in the ($t\rightarrow\infty$)-limit.


\section{The first order canonical formalism }

\ \ \ In this section, we study a canonical formulation of the first order
formalism of the Einstein gravity with  positive cosmological constant,
which is equivalent to the $SO(3,1)$ Chern-Simons gauge theory \cite{witte}.
After briefly reviewing the formalism in the general situation in \S\S 3.1,
we make a bit detailed investigation of the case $\Sigma\approx T^{2}$
and explicitly construct the spacetime in \S\S 3.2.
The resulting spacetime proves to be of the same form
as in the ADM formalism.

\subsection{The general formalism}

\ \ \ In the first order formalism, we use as independent variables the
triad one-form $e^{a}=e_{\mu}^{a}dx^{\mu}$ and the spin connection
$\omega^{ab}=\omega_{\mu}^{ab}dx^{\mu}$.
The action with the positive cosmological constant $ \Lambda $
is written as
\begin{eqnarray}
I_{CSG} &=& \int_{M}\epsilon_{abc}e^{a}\wedge
        [d\omega^{bc}+\omega^{b}_{\mbox{ }d}\wedge\omega^{dc}-
		\frac{1}{3}\Lambda e^{b}\wedge e^{c}]
		\nonumber \\*
		&=& \frac{-1}{\sqrt{\Lambda}}\int_{M}d^{3}x
\epsilon^{\mu\nu\rho}
		E_{\mu}^{a}(\partial_{\nu}\omega_{\rho
a}-\partial_{\rho}\omega_{\nu a}
		+\epsilon_{abc}\omega_{\nu}^{b}\omega_{\rho}^{c}
		-\frac{1}{3}\epsilon_{abc}E_{\nu}^{b}E_{\rho}^{c}),
\end{eqnarray}
where we have introduced new variables $E^{a}\equiv\sqrt{\Lambda}e^{a}$ and
$\omega^{a}\equiv\frac{1}{2}\epsilon^{a}_{\mbox{ }bc}\omega^{bc}$.
Let us assume $M\approx R\times\Sigma$ and construct a canonical formalism.
If we naively set $x^{0}=t$, we obtain the canonical formulation a la Witten:
\begin{equation}
I_{W}=(I_{CSG})_{|M\approx R\times\Sigma}
     =\frac{1}{\sqrt{\Lambda}}\int dt\int_{\Sigma}d^{2}x
	 (-2\epsilon^{ij}E_{ia}\dot{\omega}_{j}^{a}+E_{t}^{a}\Psi_{a}+
	                            \omega_{ta}{\cal G}^{a}),
						\label{eq:CSac}
\end{equation}
where
\begin{eqnarray}
 \Psi^{a} &=&\epsilon^{ij}(\partial_{i}\omega_{j}^{a}-
 \partial_{j}\omega_{i}^{a}+
     \epsilon^{a}_{\mbox{ }bc}\omega_{i}^{b}\omega_{j}^{c}
	            -\epsilon^{a}_{\mbox{ }bc}E_{i}^{b}E_{j}^{c}),
				\nonumber \\*
{\cal G}^{a} &=& 2\epsilon^{ij}(\partial_{i}E_{j}^{a}
                    +\epsilon^{a}_{\mbox{ }bc}\omega_{i}^{b}E_{j}^{c})
\label{eq:constr}
\end{eqnarray}
are the constraints which are of the 1st class.
 From the action (\ref{eq:CSac}) we can derive the Poisson bracket of
the basic canonical variables $(\omega_{i}^{a},E_{j}^{b})$:
\begin{equation}
\{\omega_{i}^{a}(x),E_{j}^{b}(y)\}_{P.B.}
    =\frac{\sqrt{\Lambda}}{2}\eta^{ab}\epsilon_{ij}\delta^{2}(x,y).
	\label{eq:CSGPB}
\end{equation}

Let us reduce the original phase space to obtain the physical phase space.
It is convenient to introduce a $SO(3,1)$ connection
\begin{equation}
A\equiv\omega^{a}J_{a}+E^{a}P_{a},
\end{equation}
where $(J_{a},P_{a})$ are $SO(3,1)$ generators and satisfy the
commutation relations
$$
[J_{a},J_{b}]=\epsilon_{abc}J^{c},\quad
[J_{a},P_{b}]=\epsilon_{abc}P^{c}, \quad
[P_{a},P_{b}]=-\epsilon_{abc}J^{c}.
$$
The constraint equations $\Psi^{a}\approx{\cal G}^{a}\approx 0$ tell us that
the $SO(3,1)$ connection $A$ be flat on $\Sigma$, and canonical
transformations generated by (\ref{eq:constr}) represent $SO(3,1)$ gauge
transformations. Thus we conclude that the physical phase space ${\cal M}$
of the CSG with positive cosmological constant is the moduli space of flat
$SO(3,1)$ connections modulo $SO(3,1)$ gauge transformations \cite{witte}.

To parametrize ${\cal M}$, it is convenient to use holonomies of the
connection $A$ \cite{NRZ}:
\begin{equation}
h_{A}(\gamma)\equiv {\cal P}\exp\{\int_{1}^{0}ds\dot{\gamma}^{i}(s)
                                            A_{i}(\gamma(s))\},
\label{eq:holo}
\end{equation}
where $\gamma:[0,1]\rightarrow\Sigma$ is an arbitrary closed curve on $\Sigma$
and the base point $x_{0}=\gamma(0)=\gamma(1)$ is assumed to be fixed. The
${\cal P}$ denotes the path ordered product,
with larger $s$ to the left.\\
Let us consider expressing the phase space ${\cal M}$ in terms of
(\ref{eq:holo}). Because the connection $A$ in ${\cal M}$ is flat,
the $h_{A}$ depends only on the homotopy classes of closed curve $\gamma$'s.
A gauge transformation of $A$
$$
A_{i}(x)\rightarrow A^{\prime}_{i}(x)=g(x)A_{i}(x)g^{-1}(x)
-\partial_{i}g(x)g^{-1}(x),\quad  g(x)\in SO(3,1)
$$
induces a conjugate transformation of $h_{A}$:
\begin{equation}
h_{A}\rightarrow h_{A^{\prime}}=g(x_{0})h_{A}g^{-1}(x_{0}).\label{eq:conju}
\end{equation}
Hence we can express the physical phase space as
\begin{equation}
{\cal M}=Hom(\pi_{1}(\Sigma),SO(3,1))/\sim, \label{eq:CSGPS}
\end{equation}
where $Hom(A,B)$ denotes the space of group homomorphisms :$A\rightarrow B$,
$\pi_{1}(\Sigma)$ is the fundamental group of $\Sigma$, and $\sim$ means
the equivalence under the $SO(3,1)$ conjugations (\ref{eq:conju}).

Using this parametrization of the phase space ${\cal M}$, we can compute
the maximal dimension of ${\cal M}$ in the $g\geq 2$ case to be $12g-12$,
which is the same
as that of the reduced phase space of the ADM formalism i.e. the cotangent
bundle of the Teichm\"{u}ller space \cite{witte}.

To look into particular problems such as algebras of the physical observables
and the construction of the spacetime, concrete representations of the
$SO(3,1)$ generators are required. Among them the $SO(3,1)$-vector
representation and the $SL(2,{\bf C})$-spinor representation are the most
familiar\cite{NRZ}. Their brief review is given in Appendix A.

\subsection{The torus case }

\ \ \ \ By using the general formalism explained in the previous subsection,
here we investigate geometrical
aspects of the CSG in the case of $\Sigma\approx T^{2}$.

The fundamental group $\pi_{1}(T^{2})$ of a torus is generated by two
commuting generators $\alpha$ and $\beta$. The holonomies of the flat
connection $A$ therefore form a subgroup of $SO(3,1)$ generated by two
commuting holonomies, whose general form in the self-dual representation
( see Appendix A.2 ) is
\begin{equation}\left. \begin{array}{l}
S[\alpha]\equiv S^{+}_{A}[\alpha](1,0)=\exp(\frac{\sigma_{\alpha}}{2i}
\theta^{\alpha}) \\
S[\beta]\equiv S^{+}_{A}[\beta](1,0)=\exp(\frac{\sigma_{\alpha}}{2i}
\tau^{\alpha}) \end{array} \right\}
\quad with \quad \theta^{\alpha},\tau^{\alpha}\in {\bf C}.
\end{equation}
We can easily see that the proportionality
\begin{equation}
\tau^{\alpha}=\kappa\theta^{\alpha}\qquad  with \quad \kappa\in{\bf C}
\label{eq:commute}
\end{equation}
is a sufficient condition for the holonomies to commute\footnote{
Eq.(\ref{eq:commute}) is indeed the necessary condition of the commutativity
of holonomies unless we consider the covering space of the de Sitter group
$SO(3,1)$. We do not know, however, whether it remains to be necessary or not
even when we introduce the covering space.}.
Under such a condition, we can reduce the holonomies by using appropriate
conjugations. It turns out that there are two cases:
\begin{equation}\left.  \begin{array}{ll}
\mbox{case i) \ } & S[\alpha]=\exp(\frac{\sigma_{1}}{2i}(u+i\alpha)) \\
                  & S[\beta]=\exp(\frac{\sigma_{1}}{2i}(v+i\beta))
		\end{array} \right\}\quad with \quad u,v,\alpha,\beta\in{\bf
R},
  \label{eq:RPSCS}
\end{equation}
and
\begin{equation}\left.  \begin{array}{ll}
\mbox{case ii) \ } & S[\alpha]=\exp(\frac{1}{2i}(\sigma_{2}+i\sigma_{1})) \\
                   &                                                     \\
                   & S[\beta]=\exp(\frac{k+il}{2i}(\sigma_{2}+i\sigma_{1}))
       \end{array} \right\} \quad with \quad k,l \in{\bf R}.
	 \label{eq:CSNS}
\end{equation}
Case ii) represents null transformations and we will call this sector of
the phase space the "null sector ${\cal M}_{N}$". Its detailed study
is in \S 5.

In case i), the holonomies are parametrized by four gauge-invariant
real variables. This fact shows that the physical phase space ${\cal M}$ in
the CSG on $T^{2}$ is 4-dimensional as in the ADM formalism. Let us call
this sector of the phase space the "standard sector ${\cal M}_{S}$". \\
One can easily see that the holonomies
(\ref{eq:RPSCS}) are related through the equation (\ref{eq:S-V2}) with the
$SO(3,1)$ transformations $E_{1}$ and $E_{2}$ which appeared in
eq.(\ref{eq:SO31}) in \S\S 2.2.

For the topology of the standard sector ${\cal M}_{S}$. we have two choices:
\begin{enumerate}
 \item We identify $u+2\pi$ with $u$, and $v+2\pi$ with $v$. Then the topology
  of ${\cal M}_{S}$ becomes $({\bf R}^{2}\times T^{2})/{\bf Z}_{2}$.\footnote
  {We have divided by ${\bf Z}_{2}$ because $(-u,-v;-\alpha,-\beta)$ can be
  obtained from $(u,v,\alpha,\beta)$ by a $\pi$-rotation around e.g. the
  $x^{2}$-axis. We will call this symmetry operation the "inversion".}
 \item We distinguish $(u,v)$ from $(u+2n\pi,v+2m\pi)$ ($n$ and $m$ are
 integers not being zero simultaneously).
 Then we have ${\cal M}_{S}\approx{\bf R}^{4}/{\bf Z}_{2}$.
\end{enumerate}
 From the parametrization of the holonomies (\ref{eq:RPSCS}), it seems to be
natural to choose the former. There appears, however,
to be no reason for forbidding
the use of the universal covering of the $SO(3,1)$ group since it
is sufficient for
the $SO(3,1)$ algebras to hold locally. We should choose what is more
physically relevant.

Now we extract geometrical information from the standard sector
${\cal M}_{S}$. Since the Hamiltonian vanishes in the reduced CSG, we expect
that each point of the phase space corresponds to a universe.
Witten proposes how to construct the spacetime from the phase space
\cite{witte} \cite{carli}:\it \\
 "Choose a fundamental region ${\cal F}$. It is  a part of (a covering space
of) the de Sitter space $dS^{3}$ on which the subgroup $G$ of (the covering
space of) $SO(3,1)$ generated by the holonomies of the connection $A$
properly acts. Then the quotient space:
\begin{equation}
M={\cal F}/G \label{eq:CSST}
\end{equation}
is the spacetime with $G$ as its holonomy group."\rm

We apply this prescription to our case. Since $G$ is generated by $E_{1}$ and
$E_{2}$ in (\ref{eq:SO31}) which involve pure rotation parts, ${\cal F}$
should be the universal covering of the part of $dS^{3}$ given by
(\ref{eq:embed}). Then the "angle-parameter" $\varphi$ ranges from $-\infty$
to $\infty$ and it seems more reasonable to choose the second possibility
for the topology of ${\cal M}_{S}$. The spacetime
(\ref{eq:CSST}) which we obtain from this choice takes exactly the
same form as that constructed in the ADM. This fact gives another support to
this possibility.

The problem would be that we cannot distinguish $u$ and $u+2\pi$ in the
usual representation such as (\ref{eq:RPSCS}) and (\ref{eq:SO31}). It is
possible to solve the problem formally. For instance, as Carlip has done in
presence of point particles \cite{carl2}, we replace
$\frac{\sigma_{\alpha}}{2i}$ by the
angular momentum operator $\hat{J}_{\alpha}$ and make the "wave functions"
on which the operator acts to be multivalued. Thanks to the
multivaluedness, we can
execute the desired distinction. The difference of our case from Carlip's
is that the meanings of the angular momentum operator and the "wavefunctions"
are obscure in our case. The representation which is closely related to the
spacetime structure and can naturally represent the covering space is longed
for.

We have seen Witten's construction. Alternatively, we can construct a
spacetime by finding a connection which reproduce the holonomies and
making a metric from it. Let us investigate this explicitly.\\
The simplest connection which gives the holonomies(\ref{eq:RPSCS}) is
\begin{equation}
\omega^{2}=-(\alpha dx+\beta dy)\ ,\ E^{2}=-(udx+vdy),
\qquad zero\mbox{ } otherwise,
\end{equation}
where $(x,y)$ are the coordinates on $T^{2}$ with period 1. Since this
connection gives a singular metric, we make a gauge transformation to it
by using $S_{t}\equiv \exp(t\frac{\sigma_{3}}{2})$ as a transformation matrix.
The result is:
\begin{equation}
\left\{ \begin{array}{lll} E^{\prime\mbox{ }0} &=& dt \\
                           E^{\prime\mbox{ }1} &=& -\sinh t\mbox{ }d\theta \\
						   E^{\prime\mbox{ }2} &=&
-\cosh t\mbox{ }d\varphi
	   \end{array}\right. \mbox{ , \ }
\left\{ \begin{array}{lll} \omega^{\prime\mbox{ }0} &=& 0  \\
                       \omega^{\prime\mbox{ }1} &=& \sinh t\mbox{ }d\varphi \\
				   \omega^{\prime\mbox{ }2} &=& -\cosh t\mbox{
}d\theta,
	   \end{array}\right.
\end{equation}
where as in \S\S 2.2,
$$
\varphi\equiv ux+vy\mbox{ ,} \qquad \theta\equiv \alpha x+\beta y.
$$
The spacetime whose metric is given by
\begin{equation}
ds^{2}=\frac{1}{\Lambda}[-(E^{\prime\mbox{ }0})^{2}+(E^{\prime\mbox{ }1})^{2}
                         +(E^{\prime\mbox{ }2})^{2}]
\end{equation}
is exactly the same as the spacetime determined by (\ref{eq:metric}) and
(\ref{eq:period}). This gives an alternative proof of the equivalence between
the reduced ADM formalism and the standard sector ${\cal M}_{S}$ of the CSG.


\section{Relations between the Chern-Simons gravity and the ADM formalism}

\ \ \ In the preceding sections we have seen that the standard sector
${\cal M}_{S}$ of the CSG on a torus is classically equivalent to the reduced
ADM formalism. Here we will investigate the relations between them
in more detail.

We consider first the canonical variables of the ADM
$(m_{1},m_{2},p_{1},p_{2})$ and those of the CSG $(u,v,\alpha,\beta)$.
The required relation is given by eq.(\ref{eq:phase space}) if we identify the
period parameters $(u,v,\alpha,\beta)$ in the ADM
with the canonical variables of the CSG.
For convenience we will rescale the parameters that stand for rotations
\begin{equation}
(\tilde{u},\tilde{v})\equiv\frac{1}{\sqrt{\Lambda}}(u,v),
\end{equation}
and define a new time coordinate
\begin{equation}
\tilde{\tau}\equiv\sqrt{\Lambda}\coth t.\label{eq:trep}
\end{equation}
Then, these relations are rewritten as
\begin{eqnarray}
m &\equiv& m_{1}+im_{2}=\frac{\tilde{v}+i\beta/\tilde{\tau}}
    {\tilde{u}+i\alpha/\tilde{\tau}}, \nonumber \\*
p &\equiv& p_{1}+ip_{2}=-i\tilde{\tau}(\tilde{u}-
     i\frac{\alpha}{\tilde{\tau}})^{2},
                                      \nonumber \\*
H &=& \frac{\frac{d}{dt}\tilde{\tau}}{\tilde{\tau}}
       (\tilde{u}\beta-\alpha\tilde{v})
	\equiv\frac{d\tilde{\tau}}{dt}\tilde{H}. \label{eq:C-rel}
\end{eqnarray}
These are exactly the same relations that hold in the $\Lambda=0$ case
\cite{carli}
\footnote{The correspondence between the canonical variables in these two
cases is: $$ (\tilde{u},\tilde{v},\alpha,\beta)\leftrightarrow
(a,b,\lambda,\mu). $$}.
The only difference is the range of the time variables; while
the range of $\tau$ in the $\Lambda=0$ case is $(0,\infty)$, that of
$\tilde{\tau}$ used here is restricted to $(\sqrt{\Lambda},\infty)$.
This difference seems crucial at least when we discuss trajectories.
 From these expressions, the canonical transformation from the CSG to the ADM
is easily obtained as
\begin{equation}
p_{1}dm_{1}+p_{2}dm_{2}-Hdt=2\tilde{v}d\alpha-2\tilde{u}d\beta+dF,
\end{equation}
where
$$
F(m_{1},m_{2},\alpha,\beta;\tilde{\tau})
  =\frac{1}{m_{2}\tilde{\tau}}|\beta-m\alpha|^{2}
$$
is the time-dependent generating function. The elemental Poisson brackets
in the CSG are extracted from the canonical transformation:
\begin{equation}
\{\alpha,\tilde{v}\}_{P.B.}=-\{\beta,\tilde{u}\}_{P.B.}=\frac{1}{2},
  \label{eq:CSPBT}
\end{equation}
which we could derive also from Poisson brackets of the holonomies \cite{NRZ}.
Finally we give the Hamilton equations of a physical
quantity $\tilde{O}$ in the ADM and in the CSG:
\begin{equation}\begin{array}{llc}
\frac{d}{d\tilde{\tau}}\tilde{O} &=&
   \{\tilde{O},\tilde{H}\}_{P.B.}+\frac{\partial}{\partial\tilde{\tau}}
     \tilde{O}_{|m_{1},m_{2},p_{1},p_{2}} \\
	 &=& \frac{\partial}{\partial\tilde{\tau}}
	 \tilde{O}_{|\tilde{u},\tilde{v},\alpha,\beta},
\end{array}\end{equation}
where the symbol $_{|\cdots}$ denotes that the arguments $\cdots$ are fixed.

Let us now see the quantum relation briefly.
Formally the quantum relations are obtained by replacing the classical
relations to which the canonical variables are subject by the corresponding
operator equations. In practice, there are ambiguities e.g. in the operator
ordering and a number of quantizations are constructible according to one's
viewpoints \cite{carli}\cite{ander}\cite{hosoy}. The only convincing assertion
which we can make here is the following. Since the classical relations in the
$\Lambda>0$ case are the same as those
in the $\Lambda=0$ case, the quantization
of the former is carried out in such a way that the latter is quantized;
we have only to attend to the domain of the time $\tilde{\tau}$
\footnote{If the representation which distinguishes e.g. $u+2\pi$ from $u$ is
not found, we have to consider the representation of the quantum states
to be the space of functions
of $(u,v)\in T^{2}/{\bf Z}_{2}$. The quantum theory of the CSG thus becomes
the quantum mechanics on an orbifold with vanishing Hamiltonian,
which  differs considerably from the quantizations of the $\Lambda=0$ case.}.

We will take the viewpoint of ref.\cite{carli} and begin with
scalar wavefunctions $\chi(\alpha,\beta)$ in the CSG
\footnote{According to ref\cite{witte}, it would be more natural to choose
the representation in which quantum states are functions of
the $SO(3)$ parameters $(\tilde{u},\tilde{v})$. Our choice is possible
because we have considered the phase space to be the moduli space of flat
connections of the universal covering $\widetilde{SO(3,1)}$. }.
We assume that wavefunctions $\tilde{\chi}(m,\bar{m};\tau)$ in the ADM
are related with $\chi(\alpha,\beta)$ by the integral transformation
\begin{equation}
\chi(\alpha,\beta)=\int\int\frac{d^{2}m}{m_{2}^{\mbox{ }2}}
    <m,\bar{m};\tilde{\tau}|\alpha,\beta>\tilde{\chi}(m,\bar{m};\tilde{\tau}),
	\label{eq:inttr}
\end{equation}
with the kernel
\begin{equation}
<m,\bar{m};\tilde{\tau}|\alpha,\beta>
=\frac{\beta-m\alpha}{\pi\tilde{\tau}\sqrt{2m_{2}}}
\exp(\frac{-i}{m_{2}\tilde{\tau}}|\beta-m\alpha|^{2}).\label{eq:kernel}
\end{equation}
Then the wave functions $\tilde{\chi}(m,\bar{m};\tilde{\tau})$ are naturally
given by automorphic forms of weight 1/2 on the moduli space of the torus.
The Schr\"{o}dinger equation in the ADM is given by using the weight-$1/2$
Maass Laplacian $\lap_{1/2}$ \cite{fay}:
\begin{equation}
i\frac{\partial}{\partial\tilde{\tau}}\tilde{\chi}=\tilde{\tau}^{-1}
   (\lap_{1/2})^{1/2}\tilde{\chi}\qquad
   \bigl(\lap_{1/2}\equiv -m_{2}^{\mbox{ }2}
[\frac{\partial^{2}}{\partial m_{1}^{\mbox{ }2}}
+\frac{\partial^{2}}{\partial m_{2}^{\mbox{ }2}}]
+im_{2}\frac{\partial}{\partial m_{1}}-\frac{1}{4}\bigr).
       \label{eq:schr}
\end{equation}
The probability amplitude of the event in which the time-slice at
$\tilde{\tau}_{1}$ with a modulus $m$ will have a modulus $m^{\prime}$ at
$\tilde{\tau}_{2}$ is also given by the same form as in the $\Lambda=0$ case
\cite{ezawa}:
\begin{eqnarray}
&&<m^{\prime},\bar{m^{\prime}};\tilde{\tau}_{2}|m,\bar{m};\tilde{\tau}_{1}>
                      \nonumber \\*
&&=\frac{m_{2}m^{\prime}_{2}}{4\pi\sqrt{\tilde{\tau}_{1}\tilde{\tau}_{2}}}
\frac{(\tilde{\tau}_{1}-\tilde{\tau}_{2})(\bar{m^{\prime}}-m)}
{\{(m^{\prime}_{2})^{2}+(m_{2})^{2}-m_{2}m^{\prime}_{2}
(\frac{\tilde{\tau}_{1}}{\tilde{\tau}_{2}}+
\frac{\tilde{\tau}_{2}}{\tilde{\tau}_{1}})+(m^{\prime}_{1}-m_{1})^{2}\}^{3/2}}.
                               \label{eq:tr-am}
\end{eqnarray}


\section{Special solutions for the $T^{2}$ case }

\ \ \ We have seen in \S\S 2.1 that special solutions for $g=1$
exist when $\tau$ is time independent. In \S\S 3.2 we have obtained the
null-sector ${\cal M}_{N}$ besides the standard sector ${\cal M}_{S}$ whose
detailed investigation has been given. In this section, we see the geometry
determined by the special solutions and the null-sector.

Note first that the ADM Hamiltonian vanishes when $\tau$ is time-independent,
and hence the classical solution in this case is given by a time-independent
set of the canonical variables $(m_{1},m_{2},p_{1},p_{2})$. Recall next the
discussion in \S\S 2.1, which tells us that in the present case only
nonsingular solution of the constraint equation (\ref{eq:ellip}) is given by
\begin{equation}
\tau^{2}=4\Lambda, \qquad  p_{\alpha}=0
\end{equation}
and $\lambda$ and $N$ are restricted to constants on the time-slice. Since
$N$ and $N^{i}$ are fairly arbitrarily chosen by fixing a gauge properly,
we choose the gauge that give $ N=\frac{-1}{\sqrt{\Lambda}}$ and $N^{i}=0$.
Apparently $\lambda$ is arbitrary but it is fixed by imposing the
condition that $\tau$ should be the mean curvature:
\begin{equation}
\tau=-g^{ij}K_{ij}=2\sqrt{\Lambda}.
\end{equation}
We find
\begin{equation}
\lambda=t+\frac{1}{2}\ln(m_{2}/\Lambda),
\end{equation}
where we have properly fixed the origin of the time coordinate. The spacetime
metric is given by
\begin{equation}\begin{array}{ll}
ds^{2}=&\frac{1}{\Lambda}[-dt^{2}+e^{2t}(d\xi^{2}+d\eta^{2})]\nonumber \\*
  &\left\{  \begin{array}{l}
        \xi\equiv x+m_{1}y \\
        \eta\equiv m_{2}y, \end{array}\right.
\label{eq:nullS}\end{array}
\end{equation}
where $x$ and $y$ are the coordinates on $T^{2}$ with period 1.

As in \S\S 2.2 this spacetime can be represented by the quotient space of
the upper-right-half space of the $dS^{3}$ (Figure 3)
$$
\{ (T,X,Y,Z)|-T^{2}+X^{2}+Y^{2}+Z^{2}=\frac{1}{\Lambda},\quad  T+Z>0 \}
$$
modulo identifications:
\begin{equation}
^{t}(T,X,Y,Z)\sim E^{\prime}_{1}\cdot^{t}(T,X,Y,Z)
\sim E^{\prime}_{2}\cdot^{t}(T,X,Y,Z).
\end{equation}
The transformation matrices

\begin{eqnarray}
E^{\prime}_{1} &\equiv &\exp \left(\begin{array}{cccc}
                               0 & 1 & 0 & 0 \\
							   1 & 0 & 0 & 1 \\
							   0 & 0 & 0 & 0 \\
							   0 & -1 & 0 & 0
\end{array}\right)\mbox{ ,}
					\nonumber \\*
E^{\prime}_{2} &\equiv &\exp \left( \begin{array}{cccc}
                         0 & m_{1} & m_{2} & 0 \\
						 m_{1} & 0 & 0 & m_{1} \\
						 m_{2} & 0 & 0 & m_{2} \\
						 0 & -m_{1} & -m_{2} & 0
\end{array}\right)
\end{eqnarray}
are related to the elements (\ref{eq:CSNS}) of the null sector ${\cal M}_{N}$
through the equation (\ref{eq:S-V2}) with $k+il=m_{1}+im_{2}$.

This result suggests the equivalence between the space of the special
solutions in the ADM and the null-sector ${\cal M}_{N}$ of the CSG.
We can construct the same spacetime also by
carrying out the alternative method, in which we find a connection with
the holonomies (\ref{eq:CSNS}) and construct a metric from the triad
involved in it.
This fact gives a support to the assertion that the null-sector
${\cal M}_{N}$ is equivalent to the time-independent mean-curvature case of
the ADM. Finally we give a comment that this case is similar to the trivial
$\Lambda=0$ case, where the spacetime is  static with its time-slice
unchanged, and holonomies $\Lambda_{1}$ and $\Lambda_{2}$
represent spatial translations :
$$ \begin{array}{ll}
(T,X,Y) & \stackrel{\Lambda_{1}}\longrightarrow(T,X+a,Y+b) \\
        & \stackrel{\Lambda_{2}}\longrightarrow(T,X+c,Y+d) \end{array}
$$
which are elements of the Poincar\'{e} group $ISO(2,1)$.


\section{The $\Sigma\approx S^{2}$ case}

\ \ \ \ In contrast to the $\Lambda\leq 0$ case,
the solutions for the Hamiltonian
constraint exist also when $g=0$ if the cosmological constant is positive.
In this section we will inveatigate this case.

Since $\pi_{1}(S^{2})\cong\{e\}$, where $e$ is the unity, the physical
phase space ${\cal M}$ of the CSG is a point which is characterized by the
identity holonomy. We therefore expect that the spacetime constructed by the
CSG forms a whole de Sitter space $dS^{3}$ when $\Sigma\approx S^{2}$.

In the reduced ADM formalism for $g=0$, there are no Teichm\"{u}ller
parameters or no conjugate momenta. For the moment we fix the spatial gauge as
\begin{equation}
h_{ij}dx^{i}dx^{j}=2(d\theta^{2}+\sin^{2}\theta d\varphi^{2}) \label{eq:stmt2}
\end{equation}
and the temporal gauge as
\begin{equation}
\tau=2\sqrt{\Lambda}\tanh t.
\end{equation}
Then the solutions of eq.(\ref{eq:ellip}), which form a 3-parameter family,
give the following spatial metric:
\begin{equation}
g_{ij}(x,t)=h_{ij}(x)e^{2\lambda(x,t)}=\frac{\cosh^{2}t}{2\Lambda}
      \frac{\partial\phi_{\xi}^{k}(x)}{\partial x^{i}}
	  \frac{\partial\phi_{\xi}^{l}(x)}{\partial x^{j}}
	  h_{kl}\left(\phi_{\xi}(x)\right), \label{eq:mts2}
\end{equation}
where
$$
\phi_{\xi}:\quad x^{i}\longrightarrow \phi_{\xi}^{i}(x)=\exp
  \left(\xi^{k}(x,t)\frac{\partial}{\partial x^{k}}\right)x^{i}
$$
is a diffeomorphism which is generated by a time-dependent linear combination
of three conformal killing vectors:
\begin{eqnarray}
\xi^{i}(x,t)=\alpha(t)C_{X}^{i}(x)+\beta(t)C_{Y}^{i}(x)+\gamma(t)C_{Z}^{i}(x),
                            \nonumber \\
\left\{\begin{array}{llc}
      C_{X}^{i}(x)\frac{\partial}{\partial x^{i}}
	  &=& -\cos\theta\cos\varphi\frac{\partial}{\partial\theta}+
           \frac{\sin\varphi}{\sin\theta}\frac{\partial}{\partial\varphi}\\
      C_{Y}^{i}(x)\frac{\partial}{\partial x^{i}}
	  &=& -\cos\theta\sin\varphi\frac{\partial}{\partial\theta}-
           \frac{\cos\varphi}{\sin\theta}\frac{\partial}{\partial\varphi}\\
	  C_{Z}^{i}(x)\frac{\partial}{\partial x^{i}}
	  &=& \sin\theta\frac{\partial}{\partial\theta}\quad.
	  \end{array}\right. \label{eq:ckvs2}
\end{eqnarray}
The lapse function is determined from eq.(\ref{eq:lapseD}) with $e^{2\lambda}$
given by eq.(\ref{eq:mts2}) and with $p^{ij}=0$:
\begin{equation}
N(x)=\frac{-1}{\sqrt{\Lambda}}[1+A(t)\sin(\phi_{\xi}^{\theta})
\cos(\phi_{\xi}^{\varphi})+B(t)\sin(\phi_{\xi}^{\theta})
\sin(\phi_{\xi}^{\varphi})+C(t)\cos(\phi_{\xi}^{\theta})].
\end{equation}
While $A$, $B$ and $C$ are arbitrary functions of time in general,
we will impose the condition
\begin{equation}
A^{2}(t)+B^{2}(t)+C^{2}(t)<1  \label{eq:physrel}
\end{equation}
which requires all the points in the time-slice to evolve in the same
temporal direction.
 From the discussion in \S\S 2.1 we know that the shift vector is a
time-dependent linear combination of the six conformal killing vectors.

Using these results and renaming the coordinate $\phi_{\xi}(x)$ as $x$,
we can construct the spacetime metric:
\begin{equation}
ds^{2}=\frac{1}{\Lambda}\left[\begin{array}{c}
          -(1+A(t)\sin\theta\cos\varphi+B(t)\sin\theta\sin\varphi+
		    C(t)\cos\theta)^{2}dt^{2}     \\
		   +\frac{\cosh^{2}t}{2}h_{ij}(x)
			(dx^{i}+N^{i}(x)dt)(dx^{j}+N^{j}(x)dt)
\end{array}\right].
    \label{eq:mtds3}
\end{equation}
We have used the fact that the space of the conformal killing vectors is
closed under the Lie bracket. We can use a proper time-dependent isometry of
the metric (\ref{eq:stmt2}) and absorb the killing vector part of the shift
vector, which is reduced to the following form:
\begin{equation}
N^{i}(x,t)=a(t)C_{X}^{i}(x)+b(t)C_{Y}^{i}(x)+c(t)C_{Z}^{i}(x).
                   \label{eq:shifts2}
\end{equation}
We further impose the condition that $\tau$ gives the genuine mean curvature
which is the minus trace of the extrinsic curvature. Then we find
\begin{equation}
\bigl( a(t),b(t),c(t)\bigr)=-\tanh t\bigl( A(t),B(t),C(t)\bigr).
    \label{eq:lapshif}
\end{equation}
The metric given by eqs.(\ref{eq:mtds3}), (\ref{eq:shifts2}) and
(\ref{eq:lapshif}) is induced by the Minkowski metric
$$ -dT^{2}+dX^{2}+dY^{2}+dZ^{2} $$
together with the parametrization as follows:
\begin{eqnarray}
X^{\hat{a}}&=& \Lambda(t)^{\hat{a}}_{\mbox{ }\hat{b}}X_{0}^{\hat{b}}\qquad with
                                                               \nonumber \\*
T_{0}&=& \frac{\sinh t}{\sqrt{\Lambda}}\quad and \quad
(X_{0},Y_{0},Z_{0})=\frac{\cosh t}{\sqrt{\Lambda}}
(\sin\theta\cos\varphi,\sin\theta\sin\varphi,\cos\theta),
\end{eqnarray}
where $\Lambda(t)^{\hat{a}}_{\mbox{ }\hat{b}}$ is a time-dependent Lorentz
transformation which does not depend on spatial coordinates.

As an illustration we consider the case where $A(t)=B(t)=0$.
The time-dependent Lorentz transformation in this case is given by
$$
\left(\begin{array}{l}T \\ Z \end{array}\right) =
\left(\begin{array}{ll} \cosh\Phi(t) & \sinh\Phi(t) \\
                        \sinh\Phi(t) & \cosh\Phi(t)  \end{array} \right)
\left(\begin{array}{l}T_{0} \\ Z_{0} \end{array}\right)
\quad,\quad X=X_{0} \quad and \quad Y=Y_{0},
$$
where $\Phi(t)\equiv\int^{t}dt^{\prime}C(t^{\prime})$. The time-slice
$\Sigma$ sweeps the whole de Sitter space once and only once
owing to the condition (\ref{eq:physrel}).
The similar argument can be hold in the general case and hence we deduce
that the reduced ADM gives the complete de Sitter space {\em under the
condition (\ref{eq:physrel})} and that the ADM and the CSG are equivalent
in the $g=0$ case.


\section{The anti-de Sitter universe}

\ \ \ \ So far we have studied the de Sitter case extensively. Let us now see
the anti-de Sitter case, where the cosmological constant $\Lambda$ is
negative. Henceforth we set $\Lambda=-L^{-2}$.

First of all, we can reduce the ADM formalism as in the subsection 2.1.
We have only to be careful that the coefficient $A$ is always positive in
the Hamiltonian constraint (\ref{eq:ellip}). We know that eq.(\ref{eq:ellip})
has: i) no solution if $g=0$; and ii) a unique solution if $g\geq 1$. Thus we
are interested in the $g\geq 1$ case. Since $q$ in eq.(\ref{eq:lapseD}) is
positive, the lapse function $N$ is uniquely determined. The shift vector
$N^{i}$ can be set to zero for the same reason mentioned in \S 2.

Dynamics on the phase space in the $g=1$ case is obtained by repeating the
procedure carried out in \S\S 2.2. Most of the results obtained there hold
if we replace $1/\sqrt{\Lambda}$ and $\coth t$ with $L$ and $\cot t$
respectively. Note that in the $\Lambda<0$ case the "Robertson-Walker time"
\footnote{The "Robertson-Walker time" here stands for the time which makes the
lapse function $N$ a constant.}
$t$ has a
finite domain $(0,\frac{\pi}{2})$ and that a trajectory of the Teichm\"{u}ller
parameters $(m_{1},m_{2})$ sweeps a complete semicircle in the finite
time-interval $\Delta t=\pi/2$. The spacetime constructed from the classical
solution is given by the metric:
\begin{eqnarray}
ds^{2}= L^{2}(-dt^{2}+\cos^{2}t\mbox{ }d\varphi^{2}
+\sin^{2}t\mbox{ }d\theta^{2}) \nonumber \\*
\left\{\begin{array}{l}
      \varphi=ux+vy \\ \theta=\alpha x+\beta y,
	  \end{array}\right. \label{eq:mADS}
\end{eqnarray}
where $x$ and $y$ are coordinates on $T^{2}$ with period 1.

We can embed the obtained universe in an anti-de Sitter space $AdS^{3}$ and
hence in the 2+2 dimensional Minkowski space $M^{2+2}$(see Figure 4).
We parametrize the anti-de Sitter space as:
\begin{equation}
(T,X,Y,Z)=L(\sin t\cosh\theta,\sin t\sinh\theta,\cos t\sinh\varphi,
\cos t\cosh\varphi), \label{eq:embed2}
\end{equation}
which represents a subsurface of a pseudo-sphere
$$  T^{2}-X^{2}-Y^{2}+Z^{2}=L^{2}.  $$
The metric induced by this embedding and the 2+2 dimensional Minkowski metric
$$ -dT^{2}+dX^{2}+dY^{2}-dZ^{2}  $$
reproduces the r.h.s of eq.(\ref{eq:mADS}). The periodicity condition is
encoded in the equivalence condition under two $SO(2,2)$ transformations:
\begin{equation}
^{t}(T,X,Y,Z)\sim\tilde{E}_{1}\cdot^{t}(T,X,Y,Z)
\sim\tilde{E}_{2}\cdot^{t}(T,X,Y,Z),\label{eq:adsid}
\end{equation}
where
\begin{eqnarray}
\tilde{E}_{1}&=& \left(\begin{array}{cccc}
                \cosh\alpha & \sinh\alpha & 0 & 0 \\
				\sinh\alpha & \cosh\alpha & 0 & 0 \\
				0 & 0 & \cosh u & \sinh u \\
				0 & 0 & \sinh u & \cosh u \\
				\end{array}  \right),   \nonumber \\
\tilde{E}_{2}&=& \left(\begin{array}{cccc}
                \cosh\beta & \sinh\beta & 0 & 0 \\
				\sinh\beta & \cosh\beta & 0 & 0 \\
				0 & 0 & \cosh v & \sinh v \\
				0 & 0 & \sinh v & \cosh v \\
				\end{array}  \right). \label{eq:so22}
\end{eqnarray}

Next we consider the Chern-Simons gravity. The action is that given in
\S\S 3.1, with $\Lambda$ replaced by $-L^{-2}$. We introduce an anti-de
Sitter connection:
$$ A\equiv\omega^{a}J_{a}+E^{a}\tilde{P}_{a}, $$
where $E^{a}\equiv\frac{1}{L}e^{a}$ and $(J_{a},\tilde{P}_{a})$ are $SO(2,2)$
generators subject to commutation relations:
$$
[J_{a},J_{b}]=\epsilon_{abc}J^{c}\quad , \quad
[J_{a},\tilde{P}_{b}]=\epsilon_{abc}\tilde{P}^{c}\quad , \quad
[\tilde{P}_{a},\tilde{P}_{b}]=\epsilon_{abc}J^{c}.
$$
By the discussion parallel to that in \S\S 3.1, the physical phase space
${\cal M}$ in this case turns out to be the moduli space of flat $SO(2,2)$
connections modulo $SO(2,2)$ gauge transformations.

The detailed analysis of the $\Sigma\approx T^{2}$ case can be carried out
as in \S\S3.2. We will use the spinor representation also here (see Appendix
C). Since the spinor representation is given by a
direct product of two $SL(2,{\bf R})$ representations, namely self-dual
and anti-self-dual representations $S^{(\pm)}$, each gauge transformation in
this representation is expressed by two independent local $SL(2,{\bf R})$
transformations. Taking this fact into account and by imposing the
commutativity of the holonomies around the two generators $\alpha$ and $\beta$:
$$
S^{(\pm)}[\alpha]S^{(\pm)}[\beta]=S^{(\pm)}[\beta]S^{(\pm)}[\alpha],
$$
we find nine sectors in the physical phase space ${\cal M}$. We will
denote the sector whose self-dual and anti-self-dual parts are in the $\Phi$
and $\Psi$ "subsectors" by $ {\cal M}_{(\Phi,\Psi)}$ ($\Phi,\Psi=S,N,T$).
The symbols $S$, $N$ and $T$ respectively
mean that the (anti-)self-dual part is in the spacelike subsector:
\begin{equation}
S^{(\pm)}[\alpha]=\exp\{\lambda_{2}(\alpha\pm u)\}\mbox{ , }
S^{(\pm)}[\beta]=\exp\{\lambda_{2}(\beta\pm v)\} \label{eq:SS}
\end{equation}
with $\alpha,\beta,u,v\in{\bf R}$, the null subsector:
\begin{equation}
S^{(\pm)}[\alpha]=\exp(\lambda_{0}+\lambda_{1})\mbox{ , }
S^{(\pm)}[\beta]=\exp\{(\lambda_{0}+\lambda_{1})(k\pm l)\}
\end{equation}
with $k,l\in{\bf R}$, and the timelike subsector
\begin{equation}
S^{(\pm)}[\alpha]=\exp\{\lambda_{0}(\rho\pm\phi)\}\mbox{ , }
S^{(\pm)}[\beta]=\exp\{\lambda_{0}(\sigma\pm\psi)\}
\end{equation}
with $\rho,\sigma,\phi,\psi\in{\bf R}$.
Using the relation (\ref{eq:S-V3}) between the spinor and the vector
representations, we easily see that a point of the $(S,S)$-sector given
by eq.(\ref{eq:SS}) is equivalent to a pair $(\tilde{E}_{1},\tilde{E}_{2})$
of the $SO(2,2)$ transformations in eq.(\ref{eq:so22}). Witten's construction
of the spacetime thus gives the same spacetime as the one obtained from the
classical solution of the ADM and so the $(S,S)$-sector and the phase space
of the ADM formalism are (almost) equivalent. We can show that
the classical relation of the canonical variables in these two formalisms
is given by eq.(\ref{eq:C-rel}), with $(\tilde{u},\tilde{v})\equiv L(u,v)$
and $\tilde{\tau}\equiv(\cot t)/L$. The alternative construction of the
spacetime by finding a connection with the prescribed holonomies convinces
us that it is the case.

Quantization of the $\Lambda<0$ case also could be carried out following the
same procedure as in the $\Lambda=0$ case. Because the time-slice in the
primitive space (\ref{eq:embed2}) has the topology of ${\bf R}^{2}$, we can
use as the fundamental region ${\cal F}$ the primitive space as it is.
In view of this, the $(S,S)$-sector turns out to be of the
${\bf R}^{4}/({\bf Z}_{2}\times{\bf Z}_{2})$ topology\footnote{
The two ${\bf Z}_{2}$'s in the denominator are respectively generated by the
inversion in footnote $^{7}$ and by the "mirror symmetry":
$$
(\alpha,\beta,u,v)\sim(u,v,\alpha,\beta).
$$}. We will choose the representation where wave functions are functions of
$(\alpha,\beta)$\footnote{
Precisely speaking, we should not choose such a representation since the
mirror symmetry interchanges  the coordinates with the momenta.
We could consider $(u,\alpha)$ to be coordinates on the base space
${\cal B}\approx{\bf R}^{2}/({\bf Z}_{2}\times{\bf Z}_{2})$ of the cotangent
bundle structure ${\cal M}_{(S,S)}={\bf T}^{\ast}{\cal B}$, and formally
construct the "correct" representation. The modular invariance in this
representation, however,  imposes very complicated conditions on
wave functions. We will therefore reject this possibility.}.
As is shown below, this representation gives moderate results.

The property peculiar to the anti-de Sitter case is that ${\cal M}_{(S,S)}$ in
the CSG is in 1 to 2 correspondence with the ADM phase space. The CSG has the
mirror symmetry:
\begin{equation}
(\alpha,\beta,u,v)\sim(u,v,\alpha,\beta).
\end{equation}
Using eq.(\ref{eq:C-rel}) we can rewrite this symmetry in terms of the ADM
as (Figure 5)
\begin{equation}
\tilde{\tau}\longleftrightarrow\frac{-1}{L^{2}\tilde{\tau}}
\qquad\bigl(\mbox{ }i.e.\qquad t\longleftrightarrow t-\frac{\pi}{2}\quad\bigr).
\label{eq:trev}
\end{equation}
This symmetry can be introduced also in our quantum
theory owing to the following property of the kernel:
\begin{equation}
<u,v|m,\bar{m};\tilde{\tau}>_{|(u,v)=(\alpha,\beta)}\quad=\quad
 -<\alpha,\beta|m,\bar{m};\frac{-1}{L^{2}\tilde{\tau}}>,
\end{equation}
where the l.h.s. is the Fourier transform of the kernel (\ref{eq:kernel}).
In other words, the CSG regards as gauge
equivalent the two universes which differ from each other by the replacement
(\ref{eq:trev}) and which can be distinguished in the ADM formalism.

We conclude this section by looking briefly into the geometrical aspects
of the anti-de Sitter universe. Recall that the trajectory of the modulus is
a complete semicircle as in the $\Lambda=0$ case, with the only difference
being that the range of the "Robertson-Walker time" $t$ is finite.
By noticing the fact that the
area of the time-slice is proportional to $\sin t\cos t$, we realize that
the time-slice expands from a wire-like singularity while twisting and
recollapses to another wire-like singularity.

Let us look at the mirror symmetry more closely. We see from Figure 5 that
the two gauge equivalent trajectories couples with each other to form a
circle. This implies the universe wider than that in the ADM. In the CSG,
the universe in the ADM and the "reverse universe", which is in the
lower half-space of the $AdS^{3}$ in Fig.4 and which is obtained from the
original universe by the replacement (\ref{eq:trev}), are connected through a
wire-like singularity to form a universe with three singularities.
A more radical but natural speculation proposes an "eternally oscillating
universe" obtained by the identification (\ref{eq:adsid}) of the universal
covering, which is constructed by gluing an infinite number of copies of the
anti-de Sitter space (Figure 6). We remark that these universes
cannot be constructed in the ADM.
The circle in Fig.5 necessarily traverses the $m_{1}$-axis on which the metric
$h_{ij}$ is singular and the constraint (\ref{eq:ellip}) is ill-defined. The
trajectories in the ADM are therefore restricted
within the upper (or lower) half
plane and so the ADM allows only the universes without
singularities in its middle such as shown in Figure 4.


\section{Summary and Discussion}

\ \ \ We have investigated the pure Einstein gravity with nonvanishing
cosmological constant, with main focus on the case where
the genus $g$ of $\Sigma$ is 1.
Let us summarize our main results.

The reduced ADM formalism is constructed in almost the same procedure as
in the $\Lambda=0$ case. It turns out that the Hamiltonian constraint has:
i) a unique solution if $g\geq1$; and ii) a family of (no) solutions for
$\Lambda>0(\Lambda<0)$ if $g=0$. In the case where $\Lambda>0$ and $g=0$,
it turns out that the family of solutions gives a complete de Sitter space
$dS^{3}$ under the physically relevant condition (\ref{eq:physrel}).

The physical phase space ${\cal M}$ of the CSG has been explicitly
constructed in the $g=1$ case. We have seen that for some sectors of ${\cal M}$
 the construction of the spacetime proposed by Witten works well, and (the
subspace of) the resulting spacetime proves to be identified with that
given by the solution of the ADM Hamilton equations.

We have shown that relations between phase spaces of the ADM and
the CSG is rewritten in the same expression as in the $\Lambda=0$ case,
hence dynamics on the ADM phase space ${\bf T}^{\ast}{\cal T}(T^{2})$
can be regarded to be the same. There is, in fact, additional "dynamics"
of the conformal factor fixed by the constraint, and so the dynamics of the
time-slice varies considerably according to the signature of the cosmological
constant. For $\Lambda>0$, the time-slice expands exponentially from a
wire-like singularity and its shape twists to approach a {\em nonsingular}
torus;
For $\Lambda=0$, it expands proportionally to the "Robertson-Walker time"
$t=1/\tau$ from a wire-like singularity and its shape twists to diverge to an
infinitely long band-like singularity; For $\Lambda<0$, the
time-slice expands from a wire-like singularity while twisting and then
collapses to another wire-like singularity, thus forms a closed universe.
We should bear in mind, however, that the correspondence of ${\cal M}_{(S,S)}$
to the reduced ADM phase space is 1 to 2 when $\Lambda$ is negative.
In the CSG, the spacetime
obtained by replacing $\tilde{\tau}$ by $-(L^{2}\tilde{\tau})^{-1}$ (i.e.
$t$ by $t-\pi/2$) is gauge equivalent to the original spacetime. This may
suggest the "eternally oscillating universe", which is similar to the
"tunneling solution" in the 3+1 dimensions obtained by Kodama \cite{kodama}
using the Ashtekar variables. The differences from Kodama's solution are: 1)
in our case, the tunneling into(out of) the euclidean space does not occur
because of the simultaneous changes of signs
in the $h_{ij}$ and in the $e^{2\lambda}$; and 2) our universe
is not a consequence of quantum effects but a spacetime obtained classically.

Next we discuss remaining issues. What should be elucidated are:
i) search for the representation that enables us to use $\widetilde{SO(3,1)}$
  for constructing ${\cal M}_{S}$. It is crucial because the quantum theory
  changes drastically according to its answer;
ii) construction of a genuine representation of the quantum states
for the $\Lambda<0$ case.
  Our choice of the representation is a makeshift and a
  more refined representation is needed to see,
  for example, the {\em exact} quantum relation of the CSG
  and the ADM; and
iii) to impose the modular invariance.
  There are some works\cite{carl3}\cite{hosoy} which suggest the necessity
  of taking the quotient of the Teichm\"{u}ller space modulo
   modular transformations
  $$ \gamma:(\beta,\alpha)\rightarrow(x\beta+y\alpha, z\beta+w\alpha)\quad
  \mbox{ with \ }  \left(\begin{array}{ll} x & y \\ z & w \end{array}\right)
  \in SL(2,{\bf Z}),
  $$
  where $\alpha$ and $\beta$ are two generators of $\pi_{1}(T^{2})$.
  It is thus natural to expect that wavefunctions transform covariantly
  under the modular group, it is then a nontrivial task to find such
  wavefunctions\cite{iwani}\cite{ranki}.\footnote{Note that the transition
  amplitude (\ref{eq:tr-am}) need not be invariant. Its transformation law is
  givev by
  $$
  <m',\bar{m'};\tilde{\tau}_{2}|\gamma m,\overline{\gamma m};\tilde{\tau}_{1}>
  \left(\frac{zm+w}{z\bar{m}+w}\right)^{1/2}=
  \left(\frac{-z\bar{m'}+x}{-zm'+x}\right)^{1/2}
  <\gamma^{-1}m',\overline{\gamma^{-1}m'};\tilde{\tau}_{2}|
  m,\bar{m};\tilde{\tau}_{1}>,
  $$
  where $\gamma m\equiv(xm+y)/(zm+w)$. Owing to this transformation law,
  the transformation property of weight-$\frac{1}{2}$ automorphic
  forms:
  $$ \tilde{\chi}(\gamma m,\overline{\gamma m})=
  \left(\frac{zm+w}{z\bar{m}+w}\right)^{1/2}\tilde{\chi}(m,\bar{m})
  $$
  is preserved under the evolution(\ref{eq:schr}), for a detailed discussion
  see (the revised version of) \cite{ezawa}.}

In addition, there are many interesting problems.
We know little about the sectors in the phase space ${\cal M}$ which
have no counterparts in the ADM. It is possible that fundamental properties
of the quantum gravity are hidden in these sectors. The analysis of the
null-sector ${\cal M}_{N}$ in the $\Lambda>0$ case made in this paper may
serve as a guide in the study of such sectors.
The higher genus case \cite{NRZ} is of interest since the topology-changing
processes \cite{fujiw}may be observed. Extension of the formalisms used
in this paper to the case of noncompact time-slices is expected to shed light
on the gravitational scattering of point particles in 2+1 dimensions
\cite{carl2} and on the 3 dimensional black holes \cite{henn}.

What can we say about the 3+1 dimensional quantum gravity? We cannot naively
extend the method used here to the 3+1 case. We only claim that while the
local modes are important in the 3+1 case, the global degrees of freedom are
not negligible. It would be necessary to take them into account in solving
the dynamical equations such as the Wheeler-DeWitt equation.\footnote{
The appropriate choice of the dynamical variables is of course important
\cite{carli} \cite{ashte}. }
We cannot extract immediately the useful intuition for the 3+1 case, but
we hope that our work will serve as a clue to study the more realistic
models of the quantum gravity.

\vskip2.5cm

\noindent Acknowledgments

I would like to thank Prof. K. Kikkawa, Prof. H. Itoyama and H. Kunitomo
for useful discussions and careful readings of the manuscript.


\appendix

\catcode`\@=11
\def\theequation{\Alph{section}.\arabic{equation}}

\section{The representations of the de Sitter group}

Here we give two representations for the de Sitter algebra
\begin{equation}
[J_{a},J_{b}]=\epsilon_{abc}J^{c},\quad [J_{a},P_{b}]=\epsilon_{abc}P^{c}
, \quad [P_{a},P_{b}]=-\epsilon_{abc}J^{c},
\label{eq:gener}
\end{equation}
namely, the vector and the spinor representations.

\subsection{$SO(3,1)$-vector representation}

We start from the vector representation of $SO(3,1)$ with six generators
\begin{equation}
(M_{\hat{a}\hat{b}})^{\hat{c}}_{\mbox{ }\hat{d}}=
   \delta_{\hat{a}}^{\hat{c}}\eta_{\hat{b}\hat{d}}-
       \delta_{\hat{b}}^{\hat{c}}\eta_{\hat{a}\hat{d}}
	      =-(M_{\hat{b}\hat{a}})^{\hat{c}}_{\mbox{ }\hat{d}},
\end{equation}
subject to the commutation relations:
\begin{equation}
[M_{\hat{a}\hat{b}},M_{\hat{c}\hat{d}}]=
\eta_{\hat{b}\hat{c}}M_{\hat{a}\hat{d}}-\eta_{\hat{b}\hat{d}}M_{\hat{a}\hat{c}}
+\eta_{\hat{a}\hat{d}}M_{\hat{b}\hat{c}}
-\eta_{\hat{a}\hat{c}}M_{\hat{b}\hat{d}}.
\end{equation}
If we redefine the generators as
\begin{equation}
J^{(V)}_{a}\equiv -\frac{1}{2}\epsilon_{a}^{\mbox{ }bc}M_{bc}\mbox{ , }
P^{(V)}_{a}\equiv M_{3a},
\end{equation}
then the new generators satisfy the commutation relation (\ref{eq:gener}).
In this representation, the $SO(3,1)$ connection $A$ is expressed as
\begin{equation}
(A^{(V)})^{\hat{b}}_{\mbox{ }\hat{c}}
   \equiv(\omega^{a}J^{(V)}_{a}+E^{a}P^{(V)}_{a})^{\hat{b}}_{\mbox{ }\hat{c}}
    = \left( \begin{array}{cccc}
	         0 & \omega^{2} & -\omega^{1} & -E^{0} \\
		     \omega^{2} & 0 & -\omega^{0} & -E^{1} \\
			 -\omega^{1} & \omega^{0} & 0 & -E^{2} \\
			 -E^{0}   &  E^{1}  &  E^{2}  & 0
			 \end{array} \right).
	\label{eq:V-conn}
\end{equation}
The parallel transport in terms of $A^{(V)}$ along a curve $\gamma$ from
$\gamma(t)$ to $\gamma(s)$ is expressed by the matrix
\begin{equation}
E_{A}[\gamma](s,t)^{\hat{a}}_{\mbox{ }\hat{b}}\equiv
    {\cal P}\exp\{\int_{s}^{t}du\dot{\gamma}^{i}(u)
	A^{(V)}_{i}(\gamma(u))\}^{\hat{a}}_{\mbox{ }\hat{b}},
	\label{eq:V-holo}
\end{equation}
which represents a Lorentz transformation in the 3+1 dimensional Minkowski
space. Some properties of the $E$-matrix are written down.
\begin{equation}\begin{array}{l}
\eta_{\hat{a}\hat{b}}E[\gamma]^{\hat{a}}_{\mbox{ }\hat{c}}
E[\gamma]^{\hat{b}}_{\mbox{ }\hat{d}}=\eta_{\hat{c}\hat{d}}
, \nonumber \\*
\eta_{\hat{d}\hat{c}}E[\gamma](s,t)^{\hat{c}}_{\mbox{ }\hat{b}}
\eta^{\hat{a}\hat{b}}=(E[\gamma](s,t)^{-1})^{\hat{a}}_{\mbox{ }\hat{d}}
   =E[\gamma^{-1}](t,s)^{\hat{a}}_{\mbox{ }\hat{d}}. \nonumber \\*
\epsilon_{\hat{a}\hat{b}\hat{c}\hat{d}}E[\gamma]^{\hat{a}}_{\mbox{ }\hat{m}}
  E[\gamma]^{\hat{b}}_{\mbox{ }\hat{n}}E[\gamma]^{\hat{c}}_{\mbox{ }\hat{p}}
E[\gamma]^{\hat{d}}_{\mbox{ }\hat{q}}=\epsilon_{\hat{m}\hat{n}\hat{p}\hat{q}}.
\end{array}\end{equation}

\subsection{ the $SL(2,{\bf C})$-spinor representation }

If we use as $SO(2,1)$ generators the pseudo-Pauli matrices
$$
\lambda^{\pm}_{a}\equiv(\frac{\sigma_{3}}{2i},\mp\frac{\sigma_{2}}{2},
                                       \pm\frac{\sigma_{1}}{2})
$$
which satisfy the following relations
$$
\lambda^{\pm}_{a}\lambda^{\pm}_{b}=\frac{1}{4}\eta_{ab}
                          +\frac{1}{2}\epsilon_{abc}\lambda^{\pm c},
$$
the (anti-)self-dual spinor representation of eq.(\ref{eq:gener}) is defined by
\begin{equation}
J^{\pm}_{a}\equiv\lambda^{\pm}_{a},\qquad
P^{\pm}_{a}\equiv\mp i\lambda^{\pm}_{a}.
\end{equation}
The connection $A^{\pm}$ in the (anti-)self-dual representation
is expressed as
\begin{equation}
A^{\pm}\equiv\omega^{a}J^{\pm}_{a}+E^{a}P^{\pm}_{a}=A^{(\pm)\alpha}
               \frac{\sigma_{\alpha}}{2i},
\end{equation}
where
\begin{equation}
A^{(\pm)1}\equiv E^{2}\pm i\omega^{2},
A^{(\pm)2}\equiv-(E^{1}\pm i\omega^{1}) \quad and \quad
A^{(\pm)3}\equiv \omega^{0}\mp iE^{0}.
\end{equation}
We should notice that $A^{(-)\alpha}= \overline{A^{(+)\alpha}}$, where the bar
denotes complex conjugation. The parallel transportation matrix in
the spinor representation is given by
\begin{equation}
S^{\pm}_{A}[\gamma](s,t)\equiv
    {\cal P}\exp\{\int_{s}^{t}du\dot{\gamma}^{i}(u)A^{\pm}_{i}(\gamma(u))\}.
  \label{eq:S-holo}
\end{equation}

Next we briefly summarize the spinor algebra \cite{penrose}.
The self-dual representation and the anti-self-dual representation are
respectively expressed by the unbarred indices $A,B,\cdots$ and the barred
indices $\bar{A},\bar{B},\cdots$.
We use the rank-2 antisymmetric spinor $\epsilon$ ($\epsilon^{12}=\epsilon_{12}
=1$) to lower and raise the spinor indices:
\begin{equation}
\xi_{A}=\xi^{B}\epsilon_{BA}\quad,\quad \xi^{A}=\epsilon^{AB}\xi_{B}.
\end{equation}
The similar equations hold for the barred indices. We should notice that the
results of contraction differs according to the position of indices
$$
\xi_{A}\phi^{A}=-\xi^{A}\phi_{A}.
$$
For products and traces of matrices to be well-defined, we have to fix
the positions of indices. We fix the "standard positions" of the
(anti-)self-dual representation as follows
\begin{equation}
(A^{+})_{A}^{\quad B}\mbox{ \ , \ }(A^{-})^{\bar{A}}_{\quad\bar{B}}.
\end{equation}
This choice is convenient since we can simply relate the two representations:
\begin{eqnarray}
(A^{-})^{\bar{A}}_{\quad\bar{B}}=-
\overline{\epsilon^{AC}(A^{+})_{C}^{\quad D}\epsilon_{DB}}, \nonumber \\*
S^{-}_{A}[\gamma](s,t)^{\bar{A}}_{\quad\bar{B}}=
-\overline{(S^{+}_{A}[\gamma](s,t)^{A}_{\quad B})},\nonumber \\*
{\rm tr}S^{-}_{A}[\gamma]\equiv S^{-}_{A}[\gamma](1,0)^{\bar{C}}_{\quad
\bar{C}}=  \overline{{\rm tr}S^{+}_{A}[\gamma]}.
	 \label{eq:ASD}
\end{eqnarray}
These representations have many algebraic properties,
among which the relation
\begin{equation}
^{t}\left(S^{+}_{A}[\gamma](s,t)\right)^{B}_{\quad A}=
-S^{+}_{A}[\gamma^{-1}](t,s)^{B}_{\quad A}   \label{eq:SL2C}
\end{equation}
and the similar relation for $S^{-}$ are often used.

Finally we see how the spinor representations are related to the vector
representation. If we introduce the soldering form of the $3+1$
dimensional Minkowski space:
\begin{equation}
(\sigma_{\hat{a}})_{A\bar{B}}=({\bf 1},\sigma_{1},\sigma_{2},\sigma_{3}),
\end{equation}
then the holonomy in the vector representation is reproduced from
that in the self-dual representation
\begin{equation}
S^{+}_{A}[\gamma](s,t)_{A}^{\quad C}(\sigma_{\hat{a}})_{C\bar{D}}
    (S^{+}_{A}[\gamma](s,t)^{\dagger})^{\bar{D}}_{\quad\bar{B}}
	=(\sigma_{\hat{b}})_{A\bar{B}}
	E_{A}[\gamma](s,t)^{\hat{b}}_{\mbox{ }\hat{a}},
	\label{eq:S-V}
\end{equation}
where the dagger denotes the Hermitian conjugate.
In general, the relation between the matrices of the self-dual $SL(2,{\bf C})$
representation and of the $SO(3,1)$ vector representation is the following
\begin{equation}
S\sigma_{\hat{a}}S^{\dagger}=\sigma_{\hat{b}}E^{\hat{b}}_{\mbox{ }\hat{a}},
\label{eq:S-V2}
\end{equation}
where
$$
S_{A}^{\quad B}\equiv\exp\{\frac{\sigma_{\alpha}}{2i}(\theta^{\alpha}
     +i\tilde{\theta}^{\alpha})\}_{A}^{\quad B}\quad  and \quad
E^{\hat{a}}_{\mbox{ }\hat{b}}\equiv\exp \left( \begin{array}{cccc}
          0 & \tilde{\theta}^{1} & \tilde{\theta}^{2} & \tilde{\theta}^{3} \\
		  \tilde{\theta}^{1} & 0 & -\theta^{3} & \theta^{2} \\
		  \tilde{\theta}^{2} & \theta^{3} & 0 & -\theta^{1} \\
		  \tilde{\theta}^{3} & -\theta^{2} & \theta^{1} & 0

		  \end{array} \right)
$$
are general forms of the $SL(2,{\bf C})$ and the $SO(3,1)$ matrices
respectively.


\section{the anti-de Sitter group $SO(2,2)$ }

We give two representations of the anti-de Sitter group, which are the
$SO(2,2)$-vector and the $SL(2,{\bf R})$-spinor representations.

The vector representation is given by almost the same expression as that of
$SO(3,1)$, with its metric replaced by $\eta_{\hat{a}\hat{b}}={\rm diag}
(-,+,+,-)$. We only give the explicit expression for the anti-de Sitter
connection in this representation:
\begin{equation}
(A^{(V)})^{\hat{b}}_{\mbox{ }\hat{c}}\equiv
[\omega^{a}J^{(V)}_{a}+E^{a}\tilde{P}^{(V)}_{a}]^{\hat{b}}_{\mbox{ }\hat{c}}=
\left( \begin{array}{cccc}
       0 & \omega^{2} & -\omega^{1} & E^{0} \\
	   \omega^{2} & 0 & -\omega^{0} & E^{1} \\
	   -\omega^{1} & \omega^{0} & 0 & E^{2} \\
	   -E^{0} & E^{1} & E^{2} & 0 \\
	   \end{array}\right).
\end{equation}
Its holonomy along the loop $\gamma$ is denoted by $\tilde{E}_{A}[\gamma]$.

The spinor representation is given by a direct product of two
$SL(2,{\bf R})$ representations, namely the self-dual and the
anti-self-dual representations. We express them by using pseudo-Pauli
matrices (see Appendix A.2):
\begin{equation}
J^{(\pm)}_{a}\equiv\lambda_{a}\mbox{ , }
\tilde{P}^{(\pm)}_{a}\equiv\pm\lambda_{a}.
\end{equation}
The connections in each representations are given by
\begin{equation}
A^{(\pm)}\equiv\lambda_{a}(\omega^{a}\pm E^{a}).
\end{equation}
$S_{A}^{(\pm)}[\gamma]$ denotes their holonomies.

The vector and the spinor representations are related through
the following equation:
\begin{equation}
S^{(+)}_{A}\tilde{\sigma}_{\hat{a}}(S^{(-)}_{A})^{-1}=
\tilde{\sigma}_{\hat{b}}(\tilde{E}_{A})^{\hat{b}}_{\mbox{ }\hat{a}},
\label{eq:S-V3}
\end{equation}
where we have defined the "soldering form" in 2+2 dimensional Minkowski space:
$$ \tilde{\sigma}_{\hat{a}}\equiv(2\lambda_{a}, {\bf 1}). $$
In general, the $SO(2,2)$ element which is written in the
spinor representations as
$$
S^{(\pm)}=\exp\{ \lambda_{a}(\theta^{a}\pm\kappa^{a}) \}
$$
has the following expression in the vector representation:
\begin{equation}
\tilde{E}=\exp\left( \begin{array}{cccc}
                   0 & \theta^{2} & -\theta^{1} & \kappa^{0} \\
				   \theta^{2} & 0 & -\theta^{0} & \kappa^{1} \\
				   -\theta^{1} & \theta^{0} & 0 & \kappa^{2} \\
				   -\kappa^{0} & \kappa^{1} & \kappa^{2} & 0 \\
                     \end{array} \right).
\end{equation}
These representations are, of course, subject to the relation (\ref{eq:S-V3}).


\newpage
{\bf Figure Captions}
\begin{description}
\item[Fig.1] \  The trajectory of the Teichm\"{u}ller parameters of the torus
in the $\Lambda>0$ case. \\
The trajectory forms an arc whose angle is smaller than $\pi$ (the solid line)
 when $\Lambda>0$. In the $\Lambda=0$ case it forms a semicircle centered on
 the $m_{1}$-axis (the dashed plus solid lines). The $\tau$ here denotes the
 York-time, namely the mean curvature, {\em in the $\Lambda=0$ case}.
\item[Fig.2] \ The primitive space of the torus universe in the $\Lambda>0$
case. \\
The hatched region in 2a illustrates the $Z=0$-projection of the domain where
the points (\ref{eq:embed}) ranges. The figure on the left in 2a is the
projection of the de Sitter space with $Y$- and $Z$-directions suppressed and
that on the right is the projection with $X$-direction suppressed.
Each time-slice is represented by a direct product of the hyperbola (or the
dot) on the left and the circle on the right.
\item[Fig.3] \ The fundamental region of the special solution for the torus
case. \\
Each time-slice can be expressed as a palaboloid obtained by rotating the
parabola (the solid plus dash-dotted lines) around its axis (the dotted line).
\item[Fig.4] \ The primitive space of the torus universe in the $\Lambda<0$
case.\\
The left and the right figures are respectively the projections
of the anti-de Sitter space with $Y$- and $X$-directions suppressed. The
primitive space is expressed by the shaded regions. Each time-slice is
represented by a direct product of the hyperbolae (or the dots) on the left
and on the right.
\item[Fig.5] \ The trajectory of the Teichm\"{u}ller parameters in the
$\Lambda<0$ case. \\
In the ADM the trajectory forms a complete semicircle in the upper-half plane
 (the solid line).
In the CSG we regard the trajectory in the lower-half plane (the dashed line)
as gauge equivalent to that in the upper-half plane. Here we have
deliberately used the whole complex plane so that the connectedness
of these two trajectories be obvious.
\item[Fig.6] \ Illustration of the eternally oscillating universe (the
$\alpha=v=0$ case).\\
If we take the universal covering  of the direct product space on the left,
then we obtain the universe as illustrated on the right.
\end{description}

\end{document}